\documentclass[preprint,prd,aps,showpacs,preprintnumbers,amsmath,amssymb]{revtex4-1}
\usepackage{graphics,epsfig,subfigure}
\usepackage{diagbox}
\usepackage[usenames]{color}
\usepackage{color}
\usepackage{graphicx}
\usepackage{amsfonts}
\usepackage{indentfirst}
\usepackage{booktabs}
\usepackage{hyperref}
\hypersetup{hypertex=ture,backref=true,colorlinks=ture,linkcolor=blue,anchorcolor=blue,citecolor=blue}
\usepackage{float}
\usepackage{multirow}
\usepackage[section]{placeins}
\usepackage{array}
\usepackage{amsmath}
\usepackage{slashed}

\begin{document}
\title{Dirac stars in Anti-de Sitter Spacetime}
\author{Xiao-Yu Zhang}
\author{Li-Zhao\footnote{ lizhao@lzu.edu.cn, corresponding author}}
\author{Yong-Qiang Wang\footnote{yqwang@lzu.edu.cn, corresponding author}}

\affiliation{ $^{1}$Lanzhou Center for Theoretical Physics, Key Laboratory of Theoretical Physics of Gansu Province,School of Physical Science and Technology, Lanzhou University, Lanzhou 730000, China\\
$^{2}$Institute of Theoretical Physics $\&$ Research Center of Gravitation, Lanzhou University, Lanzhou 730000, China}

\begin{abstract}
In this paper, we construct the Dirac stars model composed of two Dirac fields and Einstein gravity in four-dimensional Anti-de Sitter spacetime. We consider the Dirac stars with free field (no self-interacting).  Our investigation focuses on understanding the relationship between  Arnowitt-Deser-Misner (ADM) mass, Noether charge, and binding energy in relation to the cosmological constant. Furthermore, we extend the study to the Dirac stars with self-interacting potentials. For the self-interacting Dirac stars, three types of interactions are examined: only quartic, only sextic, quartic and sextic interactions that is kind of Q-ball type interactions. In these three distinct scenarios, it is essential to investigate the impact of self-interaction constants on Dirac stars. Additionally, we study the Dirac Q-balls in the AdS background.
\end{abstract}

\maketitle

\section{INTRODUCTION}\label{Sec1}
The idea of solitons dates back one and half centuries ago. In recent years, research on solitons has become more extensive. For the case of non-gravitational coupling, the Dirac field model is among the earliest to be researched. It can be traced back to attempts at Ivanenko \cite{Ivanenko:1938}, Weyl \cite{Weyl:1950xa}, Heisenberg \cite{Heisenberg}, and Finkelstein \cite{Finkelstein:1951zz,Finkelstein:1956hrg}. Solar has performed the first rigorous numerical solution of the Dirac solitons \cite{Soler:1970xp}. Moreover, research has been conducted on scalar field models \cite{Friedberg:1976me,Coleman:1985ki} as well as vector field models \cite{Loginov:2015rya}.

Can Dirac solitons exist when coupled with gravity? The model describes self-gravitating configurations of spin-1/2 particles, which were called the Dirac stars. In 1998, Finster et al. first studied the static spherically symmetric Dirac stars \cite{Finster:1998ws}. Since then, there have been many extensions to the study of the Dirac stars, including the coupling with the Maxwell field to form charged Dirac stars \cite{Finster:1998ux}, self-interacting Dirac stars \cite{Dzhunushaliev:2018jhj}, rotating Dirac stars \cite{Herdeiro:2019mbz}, higher energy excited Dirac stars \cite{Liang:2023ywv,Herdeiro:2017fhv} and the coupling with scalar field to form Dirac-Boson stars \cite{Liang:2022mjo}. In addition, the Dirac field has also been applied to the wormhole \cite{Blazquez-Salcedo:2020czn,Blazquez-Salcedo:2021udn,Bolokhov:2021fil,Konoplya:2021hsm,Kain:2023ann}. There have been other studies of the Dirac field coupled to gravity \cite{Daka:2019iix,Blazquez-Salcedo:2019qrz,Blazquez-Salcedo:2019uqq,Minamitsuji:2020hpl,Leith:2021urf,Dzhunushaliev:2019uft}.

To our knowledge, the studies of the Dirac stars have focused on the asymptotically Minkowski spacetime. However, inspired by the AdS/CFT correspondence \cite{Maldacena:1997re,Witten:1998qj}, researchers have extensively explored the Anti-de Sitter (AdS) spacetime. As a maximally symmetric spacetime, AdS spacetime is also a good model to study the interaction of gravity and matter field on curved backgrounds. The study of such self-gravitating systems similar to Dirac stars is extensive, with the Boson stars being the most studied. The study of Boson stars dates back to the work of Kaup \cite{Kaup:1968zz} and Ruffini \cite{Ruffini:1969qy} 50 years ago, they found a solution to the spherically symmetric Einstein-Klein-Gordon equation. Until 2003, Astefansesi and Radu have proposed the Boson stars in AdS spacetime \cite{Astefanesei:2003qy}. After this, there has been a lot of promotion of Boson stars in AdS spacetime. The self-interacting Boson stars in AdS spacetime have been studied by Hartmann et al. \cite{Hartmann:2012gw,Hartmann:2012wa}. Brihaye has studied the rotating Boson stars in AdS spacetime \cite{Brihaye:2014bqa}. The coupling of the scalar field in the Maxwell field and gravity constitutes charged Boson stars in AdS spacetime \cite{Brihaye:2013hx,Brihaye:2022oaf,Hu:2012dx,Guo:2020bqz}. There have been many interesting studies on the model of the scalar field coupled gravity in AdS spacetime \cite{Brihaye:2013tra,Radu:2012yx,Kichakova:2013sza,Nogueira:2013if,Buchel:2013uba,Liu:2020uaz}. In contrast to the AdS spacetime, there have been also studies on Boson stars in dS spacetime \cite{Hartmann:2013kna}. Extending the scalar field to the vector field can form the Proca stars in AdS spacetime \cite{Duarte:2016lig}. In our work, we construct the spherical Dirac stars in AdS spacetime, and investigate the characteristics of various solutions. Additionally, we investigate Dirac Q-balls, a configuration of non-topological solitons that exists without gravity, within the AdS background.

The organization of this paper is as follows. In Sec. \ref{sec2}, we construct a four-dimensional model in AdS spacetime involving two Dirac fields and gravity coupling. In Sec. \ref{sec3}, we provide the initial conditions and boundary conditions. Sec. \ref{sec4} shows the numerical results of the motion equations, presenting solutions for the free-field Dirac stars, the interacting Dirac stars, and the Dirac Q-balls. Finally, we make the summaries in sec. \ref{sec5}.

\section{THE MODEL SETUP}\label{sec2}
\subsection{Framework}
We consider the system of Einstein gravity to be Minimal coupled to two Dirac fields in 3+1-dimensional spacetime, and  the action is given by
    \begin{equation}\label{action}
        S=\int d^4x\sqrt{-g}\left(\frac{R-2\Lambda}{16\pi G}+\mathcal{L}_{D}\right),
    \end{equation}
where $G$ and $\Lambda$ are the Newtonian constant and the cosmological constant, respectively. $R$ is the Ricci scalar, and  $\mathcal{L}_{D}$ denotes the Lagrangians of the Dirac field, which is expressed as
    \begin{eqnarray}
        \mathcal{L}_{D}=\sum_{k=1}^{2}\mathcal{L}_{(k)} = -\sum_{k=1}^{2} \left[\frac{\mathrm{i}}{2}\left(\left\{\hat{\slashed{D}} \overline{\Psi}^{(k)}\right\}\Psi^{(k)} - \overline{\Psi}^{(k)} \hat{\slashed{D}} \Psi^{(k)} \right) + V\left(\Phi^2\right)\right],
    \end{eqnarray}
    \begin{eqnarray}
        V\left(\Phi^2\right) =  \mu \Phi^2 - \xi \Phi^4 + \nu \Phi^6.
    \end{eqnarray}
Due to the spherical symmetry constraint, we require two Dirac fields, $\overline{\Psi}^{(k)}$ are the Dirac conjugate. $V\left(\Phi^2\right)$ is the potential energy term, where $\mu$ is the mass of the Dirac field, and $\xi,\nu$ are coupling constants for the interaction. In order to simplify, we define $ \Phi^{2} = \mathrm{i}\overline{\Psi}^{(k)}\Psi^{(k)}$, then $\Phi^{4}=\left(\Phi^{2}\right)^{2}$, $\Phi^{6}=\left(\Phi^{2}\right)^{3}$. $\hat{\slashed{D}}=\gamma^{\mu}\hat{D}_{\mu}$, where $\gamma^{\mu}$ are the gamma matrices in curved spacetime. $\hat{D}_{\mu}=\partial_{\mu}-\Gamma_{\mu}$ is the spinor covariant derivative, where $\Gamma_{\mu}$ is the spinor connection matrices. The field equations are given by
    \begin{eqnarray}
        \mathcal{R}_{\alpha\beta}-\frac{1}{2}\mathcal{R} \mathrm{g}_{\alpha\beta}+\Lambda \mathrm{g}_{\alpha\beta}=8 \pi \mathrm{G} T_{\alpha\beta},
        \end{eqnarray}
        \begin{eqnarray}
        \left(\hat{\slashed{D}}-\frac{\partial V}{\partial \psi^{2} }\right)\Psi^{(k)}=0,
    \end{eqnarray}
where $T_{\alpha\beta}$ is the energy-momentum tensors of the Dirac fields
    \begin{eqnarray}
        T_{\alpha\beta}=\sum_{k=1}^{2} T^{(k)}_{\alpha\beta},\qquad T^{(k)}_{\alpha\beta} = -\frac{\mathrm{i}}{2}\left[\overline{\Psi}^{(k)}\gamma_{\left(\alpha\right.}\hat{D}_{\left.\beta\right)}\Psi^{(k)}
        -\left\{\hat{D}_{\left(\alpha\right.}\overline{\Psi}^{(k)} \right\}\gamma_{\left.\beta\right)} \Psi^{(k)} \right]-\mathrm{g}_{\alpha\beta}\mathcal{L}_{(k)}.
    \end{eqnarray}
The detailed derivation of the energy-momentum tensor is in Ref. \cite{Shapiro:2016pfm}. The action of the matter field is invariant under the $ U(1)$ transformation $\Psi^{(k)}\rightarrow e^{i\alpha}\Psi^{(k)}$, where $\alpha$ is a constant. Therefore, there exists a conserved current
    \begin{eqnarray}
        &&j^{\alpha} = \overline{\Psi}\gamma^{\alpha}\Psi,
    \end{eqnarray}
and the conserved charge can be obtained
    \begin{eqnarray}
        &&Q = \int_{\Sigma} j^{\mathrm{t}},
    \end{eqnarray}
where $ j^{\mathrm{t}}$ is the time component of the current, and $\Sigma$ is a timelike hypersurface. The magnitude of the charge represents the number of particles in the system.
\subsection{Ansatz and equation of motion}
We consider a static spherically symmetric configuration and choose the following form of the spacetime metric
    \begin{eqnarray}
        ds^{2}=-N\left(r\right)\sigma^{2}\left(r\right)dt^{2}
        +\frac{dr^2}{N\left(r\right)}
        +r^{2}\left(d \theta^{2}+\sin^{2}\theta d \varphi^{2} \right),
    \end{eqnarray}
where $N(r)=1-\frac{2m(r)}{r}-\frac{\Lambda r^{2}}{3}$, $m(r)$ and $\sigma (r)$ depend only on the radial distance $r$. For the Dirac field we use the following ansatz
    \begin{equation}
         \Psi^{(1)} = \begin{pmatrix}\cos(\frac{\theta}{2})[(1 + i)f(r) - (1 - i)g(r)]\\ i\sin(\frac{\theta}{2})[(1 - i)f(r) - (1 + i)g(r)]\\-i\cos(\frac{\theta}{2})[(1 - i)f(r) - (1 + i)g(r)]\\ -\sin(\frac{\theta}{2})[(1 + i)f(r) - (1 - i)g(r)] \end{pmatrix}e^{i\frac{\varphi}{2} - i\omega t}\,,
    \end{equation}
    \begin{equation}
         \Psi^{(2)} = \begin{pmatrix}i\sin(\frac{\theta}{2})[(1 + i)f(r) - (1 - i)g(r)]\\ \cos(\frac{\theta}{2})[(1 - i)f(r) - (1 + i)g(r)]\\ \sin(\frac{\theta}{2})[(1 - i)f(r) - (1 + i)g(r)]\\ i\cos(\frac{\theta}{2})[(1 + i)f(r) - (1 - i)g(r)] \end{pmatrix}e^{-i\frac{\varphi}{2} - i\omega t}\,,
    \end{equation}
where $f(r)$ and $g(r)$ are real functions, $\omega$ is the frequency of the Dirac field.
By substituting the ansatz into the Einstein equation and the Dirac equation, we obtain
    \begin{eqnarray}\label{12-15}
        &&m^{\prime} = 2r^{2}\left[4\sqrt{N}\left(gf^{\prime}-fg^{\prime}\right)+\frac{8fg}{r}
        + V \right],\label{eom12}   \\
        &&\frac{\sigma^{\prime}}{\sigma}= \frac{8r}{\sqrt{N}}\left(gf^{\prime}-fg^{\prime}
        +\frac{\omega\left(f^{2}+g^{2}\right)}{N\sigma}\right), \label{eom13}     \\
        &&f^{\prime}+\left(\frac{N^{\prime}}{4N}+\frac{\sigma^{\prime}}{2\sigma}+\frac{1}{r\sqrt{N}}+\frac{1}{r}\right)f
        -\frac{\omega g}{N \sigma}+\frac{g}{\sqrt{N}}\frac{\partial V}{\partial \psi^{2}} =0,         \\
        &&g^{\prime}+\left(\frac{N^{\prime}}{4N}+\frac{\sigma^{\prime}}{2\sigma}-\frac{1}{r\sqrt{N}}+\frac{1}{r}\right)f
        +\frac{\omega f}{N \sigma}+\frac{f}{\sqrt{N}}\frac{\partial V}{\partial \psi^{2}} =0.\label{eom15}
    \end{eqnarray}
This is a coincidence, the equations here are quite similar to those of asymptotically Minkowski spacetime, with the only difference being the inclusion of a cosmological constant in the $N(r)$.  The conserved charge is
    \begin{eqnarray}
        Q=8\int_{0}^{\infty}dr \, r^{2}\frac{\left(f^{2}+g^{2}\right)}{\sqrt{N}},
    \end{eqnarray}
and the energy density is $\rho=-\mathcal{T}_{t}^{t}$
    \begin{eqnarray}
        \rho=8\left(\left(gf^{\prime}-fg^{\prime}\right)\sqrt{N}+\frac{2fg}{r}+\frac{1}{4}V\right).
    \end{eqnarray}

\section{BOUNDARY CONDITIONS}\label{sec3}
If we need to obtain numerical solutions for the above equations, it is necessary to choose appropriate initial conditions and boundary conditions. The conditions that the metric functions need to satisfy at the origin and infinity are
    \begin{eqnarray}
        m\left(0 \right)=0, \quad \sigma\left(0 \right)=\sigma_{0}, \quad
        m\left(\infty \right)= M, \quad \sigma\left(\infty \right)=1,
    \end{eqnarray}
where $M$ and $\sigma\left(0\right)$ are currently unknown. The ADM mass can be obtained from the behavior of the metric function $m(r)$ at infinity. For the Dirac Q-balls, the ADM mass is obtained by integrating the energy density $T^{0}_{0}$
\begin{eqnarray}
        &&M = -\int^{\infty}_{0}dr \,  r^{2} T^ {0}_{0},
    \end{eqnarray}
and this integral is also a finite value in AdS spacetime. The conditions for the matter field at the origin and at infinity are
    \begin{eqnarray}
        f\left(0 \right)=0,  \quad  \left.\frac{dg\left(r \right)}{dr}\right|_{r=0}=0,\quad f\left(\infty \right)=g\left(\infty \right)=0.
    \end{eqnarray}
\section{NUMERICAL RESULTS}\label{sec4}
To simplify numerical computations, we use dimensionless quantities, employing two distinct transformations. In studying the free field Dirac stars and the only quartic or only sextic interacting Dirac stars, the equations depend on two potential parameters $\mu$, $\xi$ or $\nu$, as well as the Newtonian constant $G$ and the cosmological constant $\Lambda$. In this case, we adopt
    \begin{eqnarray}
        r\rightarrow r\mu,\quad\omega\rightarrow\omega/\mu,\quad \nu\rightarrow\nu \mu
        ,\quad f\rightarrow \sqrt{\frac{4 \pi G}{\mu}}f ,\quad g\rightarrow \sqrt{\frac{4 \pi G}{\mu}}g.
    \end{eqnarray}
We set units such that $\mu=1$ and $4 \pi G=1$, which is a common scaling adopted in the study of the Dirac stars. When studying the Q-ball type interacting Dirac stars and Dirac Q-balls, these equations depend on three potential parameters $\mu,\xi,\nu$, as well as the Newtonian constant $G$ and the cosmological constant $\Lambda$. We need to simplify the parameters, and adopt the second transformation
    \begin{eqnarray}
        \nu\rightarrow \frac{\nu}{\xi},\quad f\rightarrow \sqrt{\xi}f ,\quad g\rightarrow \sqrt{\xi}g,
    \end{eqnarray}
the radial coordinate and the field frequency remain unaffected. By setting units with $\mu=\xi=1$, the equations can then depend on two dimensionless parameters and the cosmological constant. We define these two dimensionless parameters as $\bar{\nu}=\nu\mu/\xi^{2}=\nu$ and $\alpha=4 \pi G / \xi=4 \pi G$, where the range of $\alpha$ is $\alpha\in[0,\infty)$. $\alpha=0$ corresponds to $G=0$, which is a solution fixed to the spacetime background, the Dirac Q-balls. When $\alpha\rightarrow\infty$, it corresponds to non-interacting Dirac stars. After setting the units, it can be observed that the two different transformations yield identical results.
In addition, we introduce a new radial variable
     \begin{eqnarray}
         x=\frac{r}{1+r},
    \end{eqnarray}
where the radial coordinate is $r\in[0,\infty)$, so $x\in[0,1]$. We employ the Newton-Raphson method as our iterative technique and numerically solve the system of differential equations using the finite element method. To ensure the accuracy of the solutions, we require that the relative error be less than $10^{-5}$. Starting from the next section, we will present our numerical results. We will discuss four different scenarios in three subsections: free-field Dirac stars (Case 1: $\xi=0$ and $\nu=0$), quartic or sextic interacting Dirac stars (Case 2: $\xi=0$ or $\nu=0$), and Q-ball type interacting Dirac stars (Case 3: $\xi=1$ and $\nu=1$).
\subsection*{Case 1: $\xi=0$ and $\nu=0$}
    \begin{figure}[b]
         \centering
         \subfigure{\includegraphics[width=0.49\textwidth]{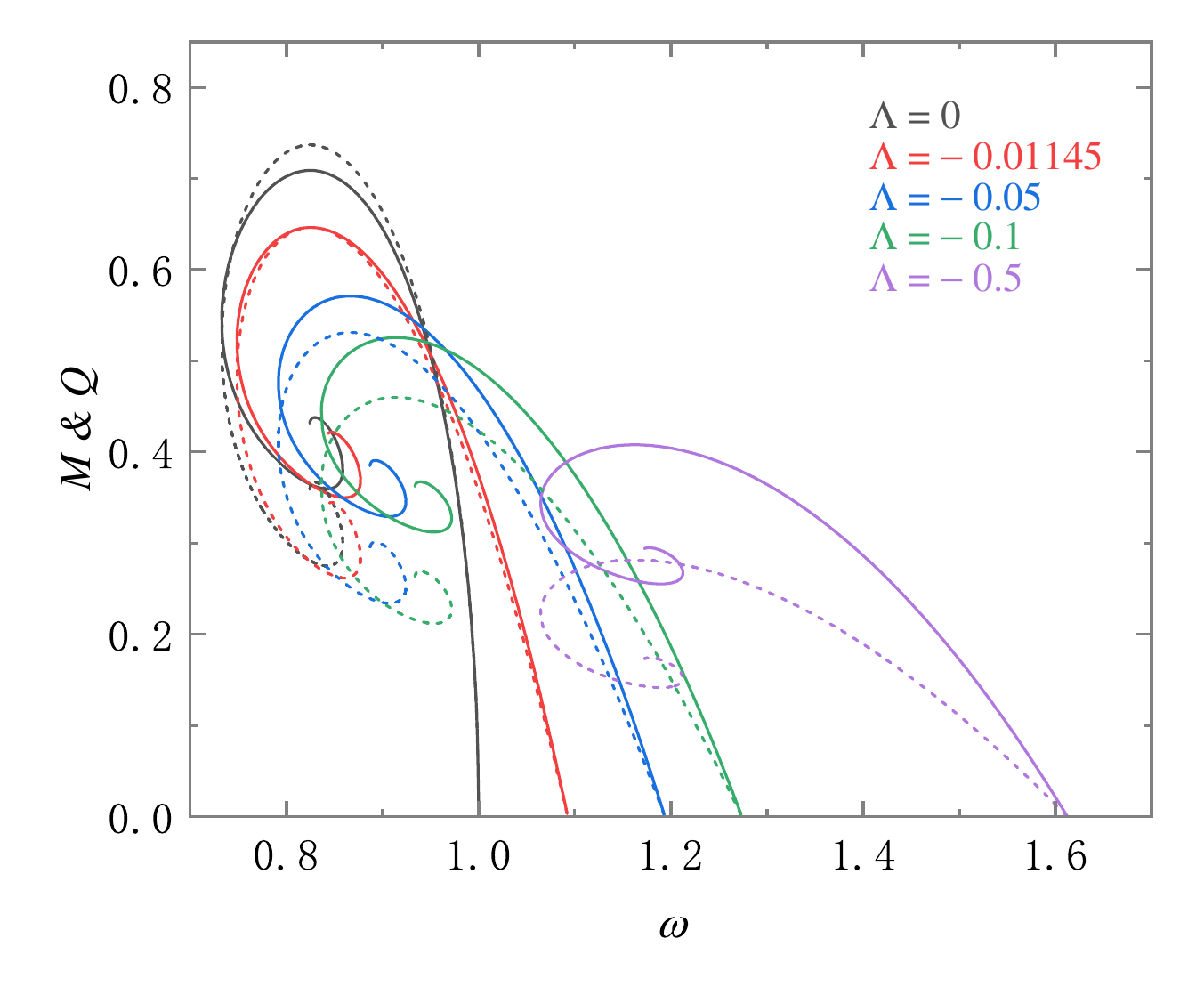}}
         \subfigure{\includegraphics[width=0.49\textwidth]{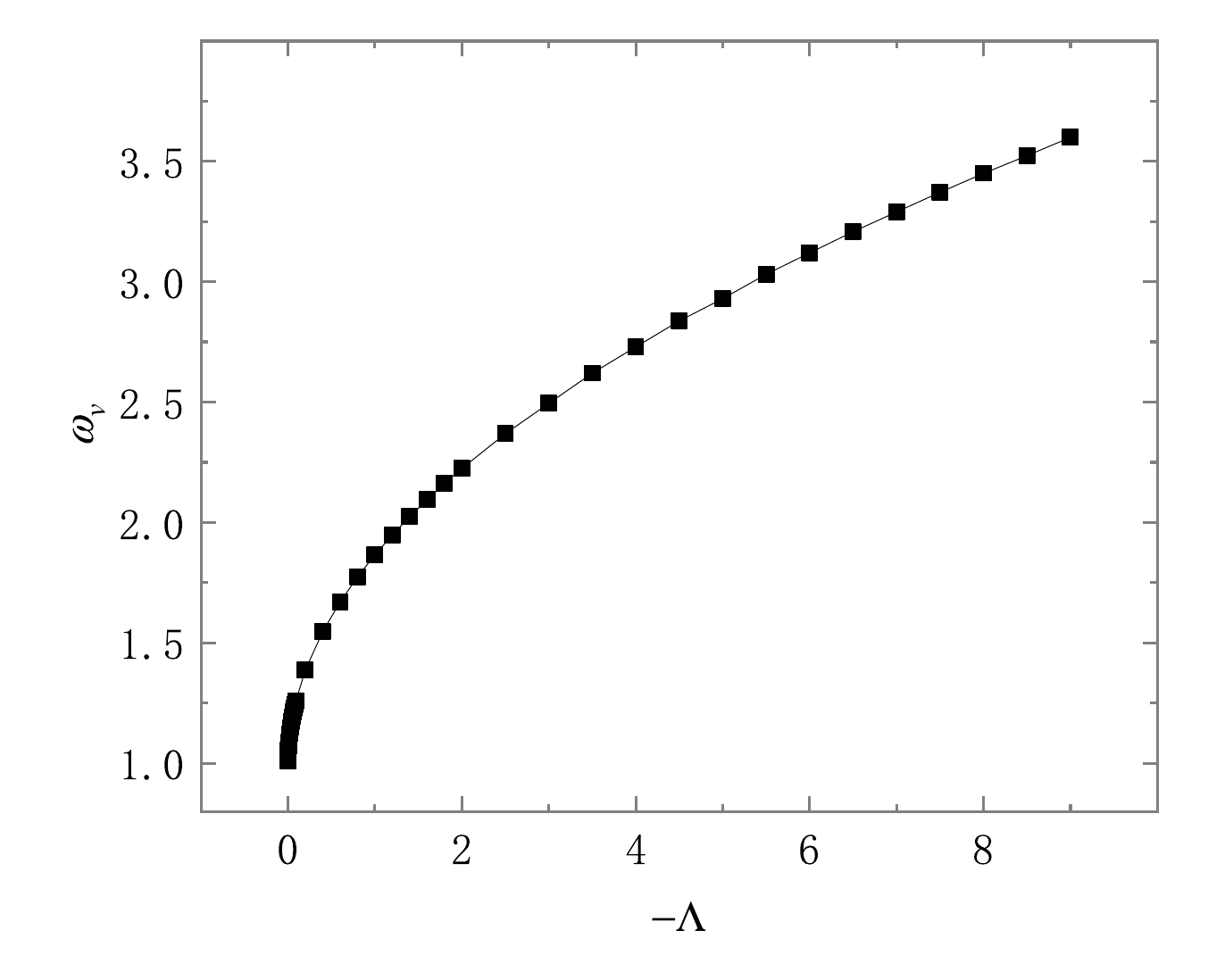}}
         \caption{Left: the mass $M$ (solid line) and charge $Q$ (dashed line) of the Dirac stars as functions of frequency $\omega$ under different cosmological constants $\Lambda$. Right: the frequency $\omega_{\textrm{v}}$ as a function of the cosmological constant $-\Lambda$.}
         \label{pic000}
    \end{figure}
First, we investigate the free field Dirac stars, meaning no self-interaction ($\xi=0$ and $\nu=0$). By solving the motion equations (\ref{eom12}-\ref{eom15}), we obtain numerical solutions satisfying the boundary conditions.

The mass $M$ and charge $Q$ of the Dirac stars as a function of frequency $\omega$ for different cosmological constants $\Lambda$ is shown in the left panel of Fig. \ref{pic000}. Similar to Dirac stars in asymptotically Minkowski spacetime, the cosmological constant does not alter the characteristics of the spirals. They exist within a frequency range $(\omega_{min},\omega_{max})$, and the mass and charge approach zero at $\omega_{\textrm{v}}$.  For the Dirac stars with free field, $\omega_{max}=\omega_{\textrm{v}}$. The variation of $\omega_{\textrm{v}}$ with the cosmological constant is depicted in the right panel of Fig. \ref{pic000}. It can be observed that, as the cosmological constant decreases, $\omega_{\textrm{v}}$ gradually increases. The position and size of the spirals are influenced by the cosmological constant, the frequency of the position increases as the cosmological constant decreases. The maximum mass $M_{max}$ of the Dirac stars decreases as the cosmological constant decreases, and a similar trend is observed for the maximum charge $Q_{max}$. Additionally, we observe that in asymptotically Minkowski spacetime, the mass and charge of the Dirac stars exhibit a steep slope during the process of increasing from zero to their maximum values. As it approaches $\omega_{\textrm{v}}$, this slope tends to be vertical. Furthermore, as the frequency decreases, the trend of changes in mass and charge becomes more gradual.

\begin{figure}[h]
         \centering
         \subfigure{\includegraphics[width=0.5\textwidth]{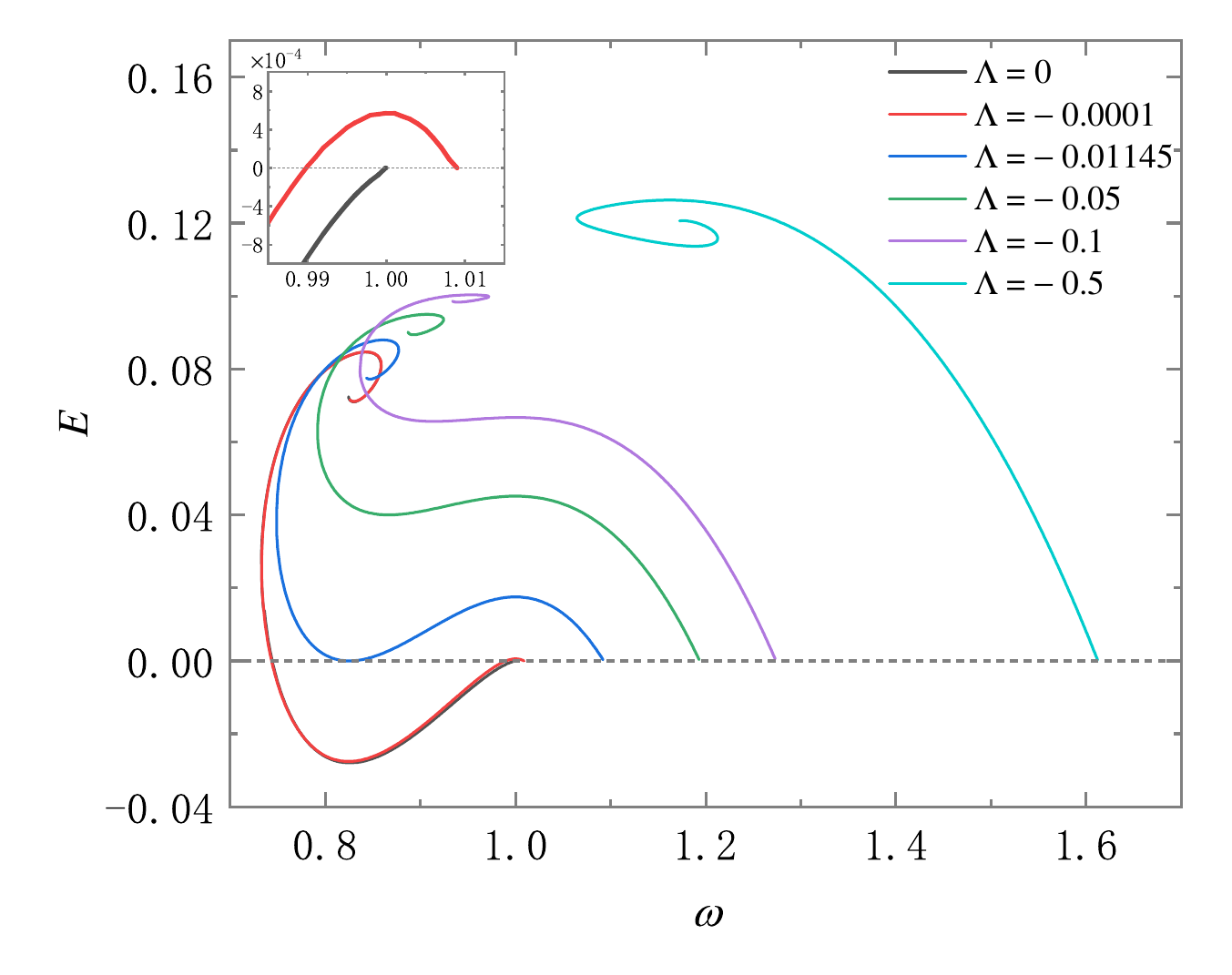}}
         \caption{The binding energy $E$ of the Dirac stars as a function of frequency $\omega$ for different cosmological constants $\Lambda$, with the inset providing a detailed view around $\omega=1$. The case $\Lambda=-0.01145$ corresponds to a critical situation where the binding energy is non-negative.}
         \label{pic000E}
    \end{figure}

The variation of the binding energy $E$ of the Dirac stars with frequency $\omega$ is illustrated in Fig. \ref{pic000E}. In asymptotically Minkowski spacetime, stable solutions $(E<0)$ emerge on the first branch of Dirac stars. However, with a decrease in the cosmological constant, these stable solutions gradually disappear. When $\Lambda=-0.01145$, the entire binding energy becomes non-negative, confirming the lack of stable solutions. The curve depicting the variation in binding energy also displays spirals, and as the cosmological constant decreases, the direction of the spirals changes. In the inset, we have magnified  the region around $\omega=1$. For the case of $\Lambda=0$, the binding energy of the Dirac stars near $\omega_{\textrm{v}}$ decreases as the frequency decreases, resulting in negative binding energies. However, when $\Lambda\neq0$, the binding energy near $\omega_ {\textrm{v}}$ can be greater than zero. For example, in the case of $\Lambda=-0.0001$, the binding energy of the Dirac stars exhibits only a small region where it exceeds zero.

\subsection*{Case 2: $\xi=0$ or $\nu=0$}
    \begin{figure}[b]
        \centering
        \subfigure{\includegraphics[width=0.49\textwidth]{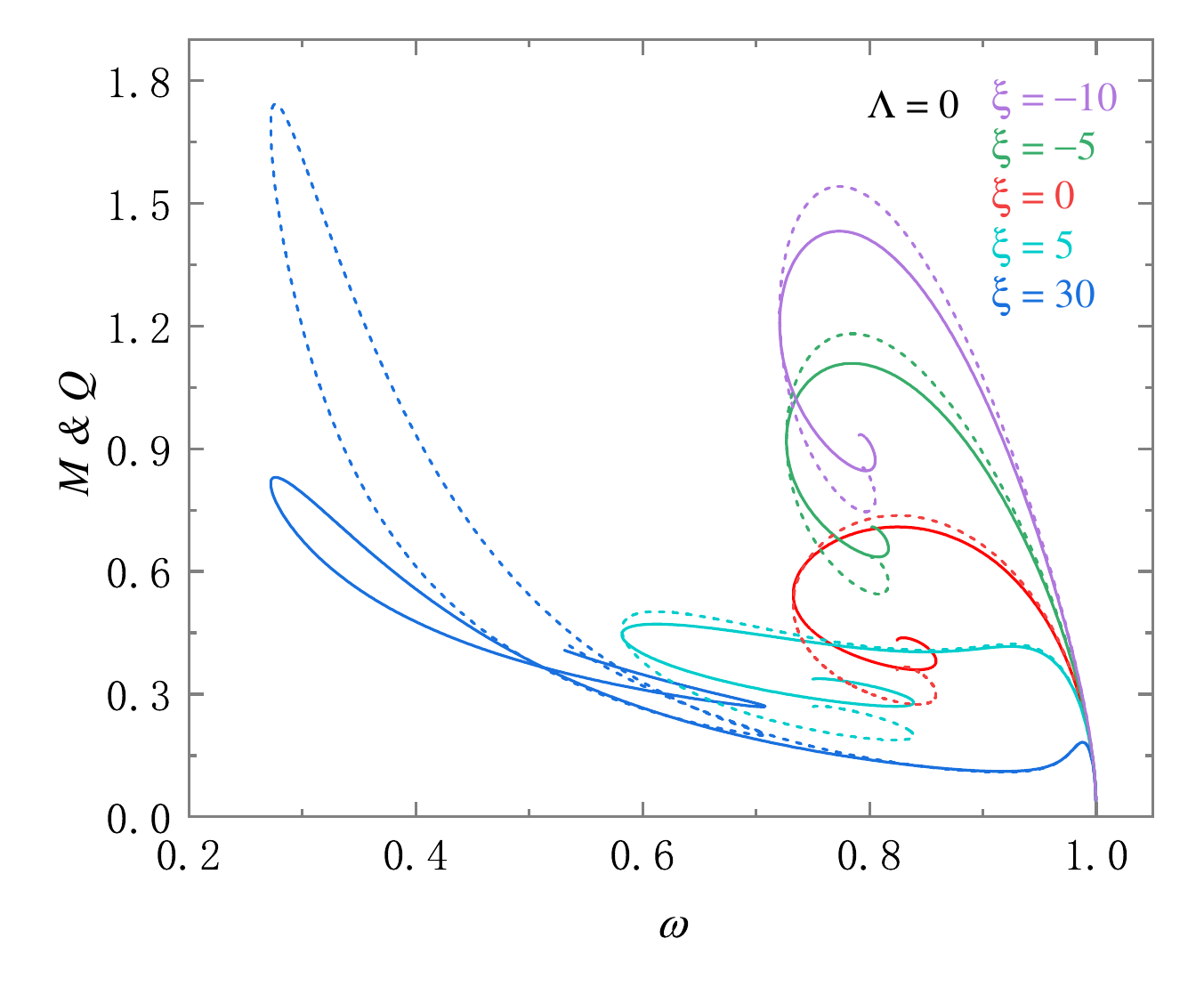}}
        \subfigure{\includegraphics[width=0.49\textwidth]{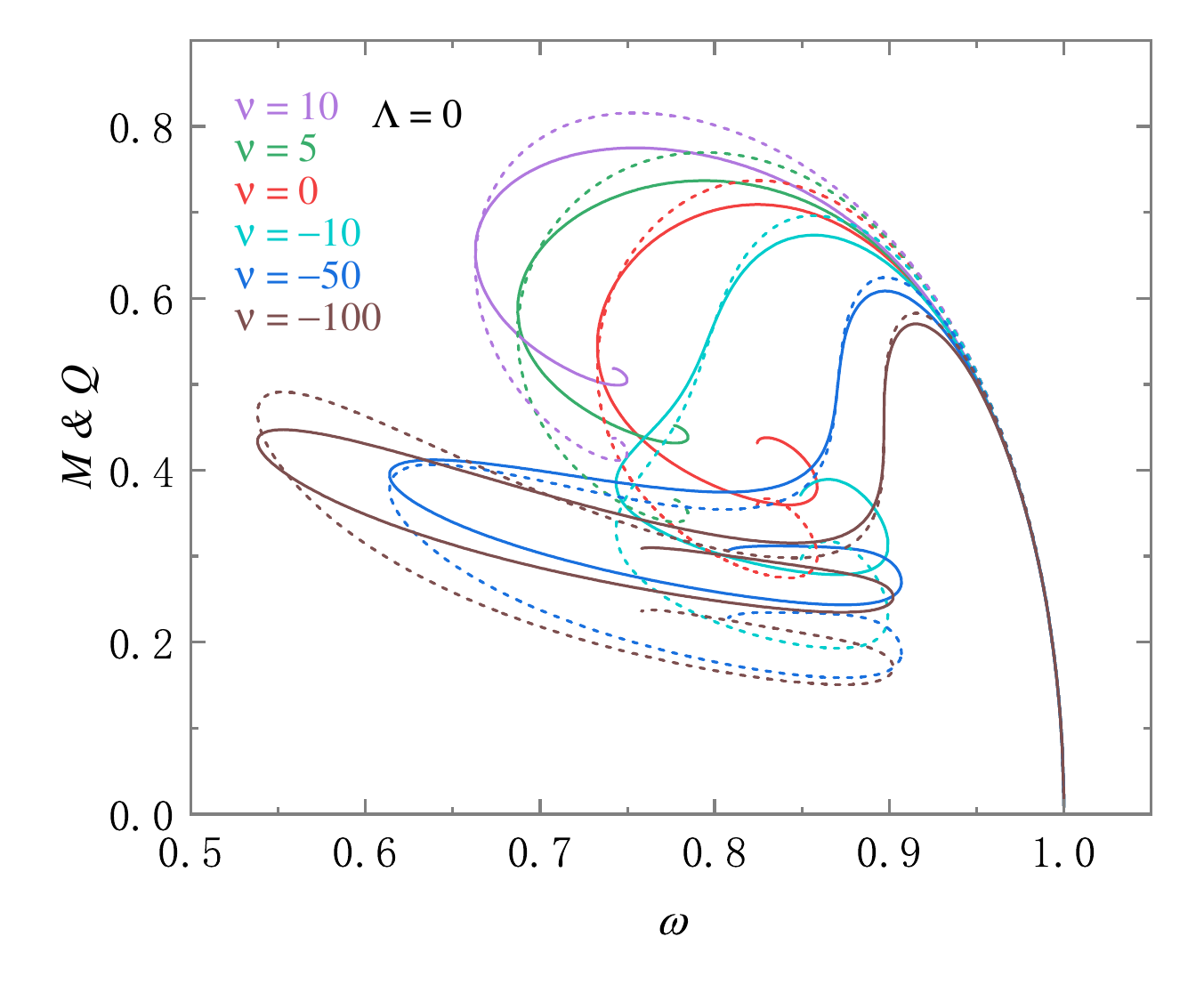}}
        \subfigure{\includegraphics[width=0.49\textwidth]{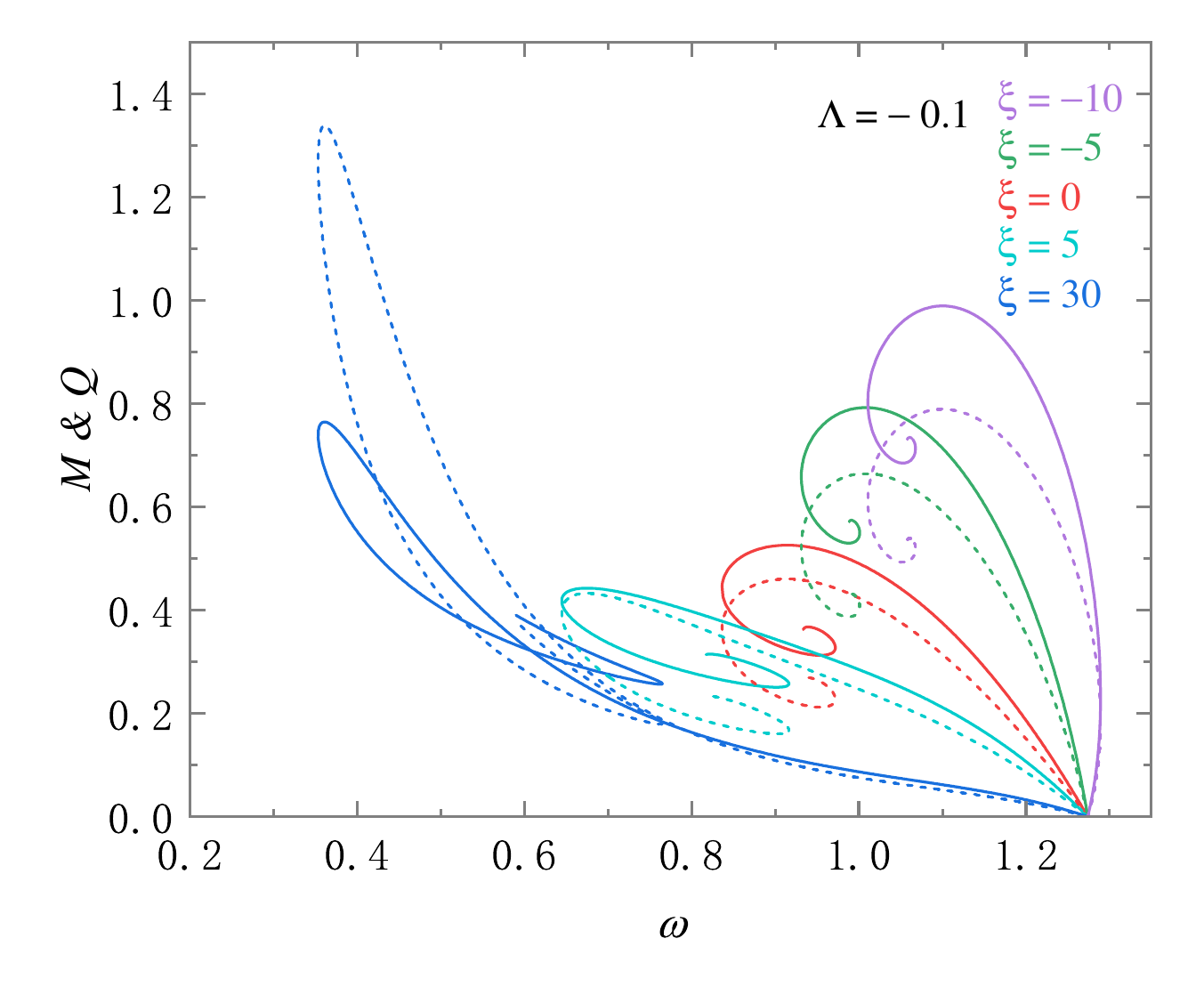}}
        \subfigure{\includegraphics[width=0.49\textwidth]{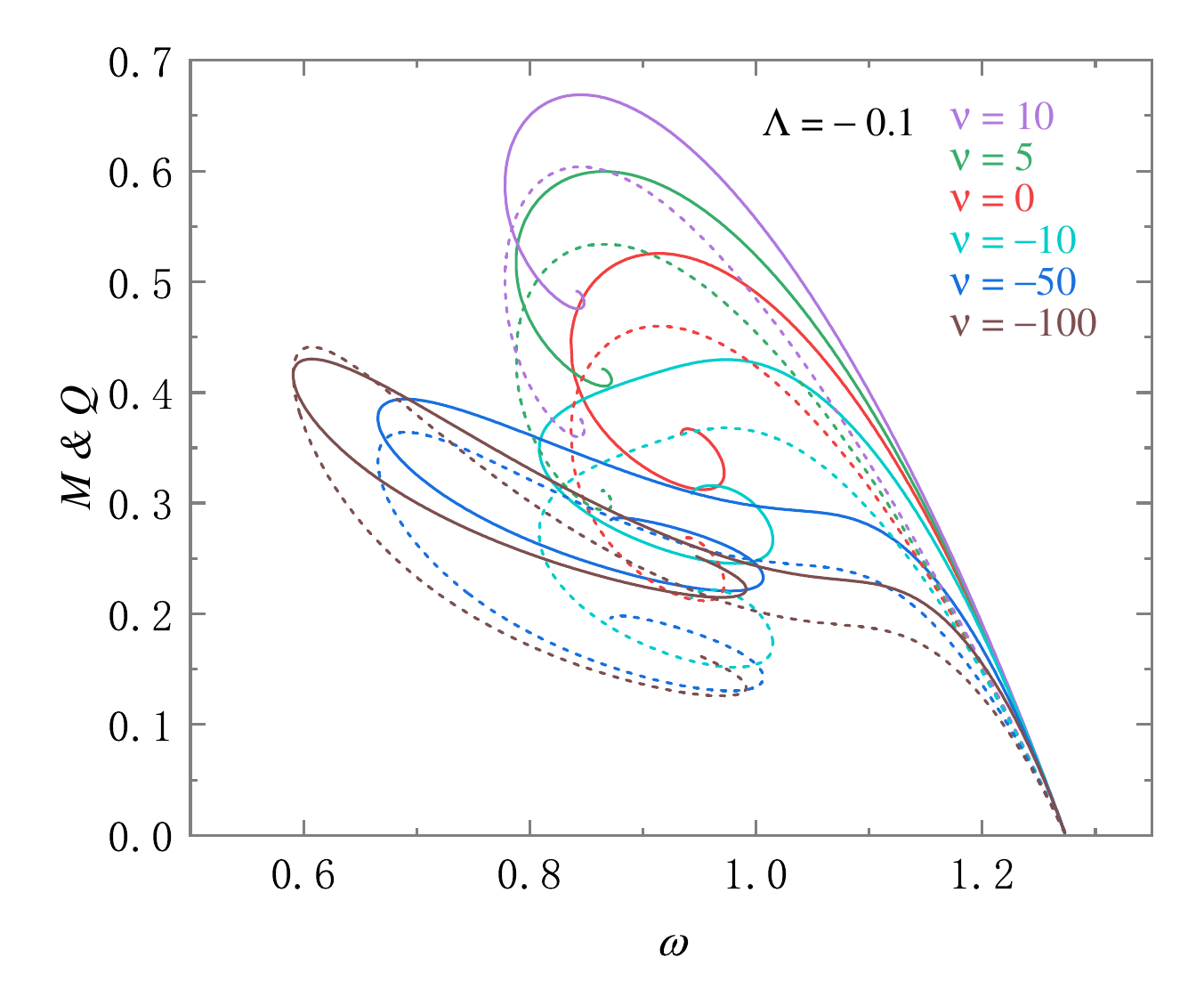}}
        \caption{The mass $M$ (solid line) and charge $Q$ (dashed line) of the Dirac stars as functions of frequency $\omega$ for different interaction parameters $\xi$ or $\nu$ in asymptotically Minkowski spacetime ($\Lambda=0$) or AdS spacetime ($\Lambda=-0.1$). The left column corresponds to $\Phi^{4}$ interactions, while the right column corresponds to $\Phi^{6}$ interactions.}
        \label{psi46MQ}
    \end{figure}
In this section, we investigate the Dirac stars with quartic or sextic interactions. To more intuitively illustrate the impact of different interactions,
the left column in all the figures corresponds to $\xi \Phi^{4}$ interactions, and the right column corresponds to $\nu \Phi^{6}$ interactions.

\begin{figure}
         \centering
         \subfigure{\includegraphics[width=0.49\textwidth]{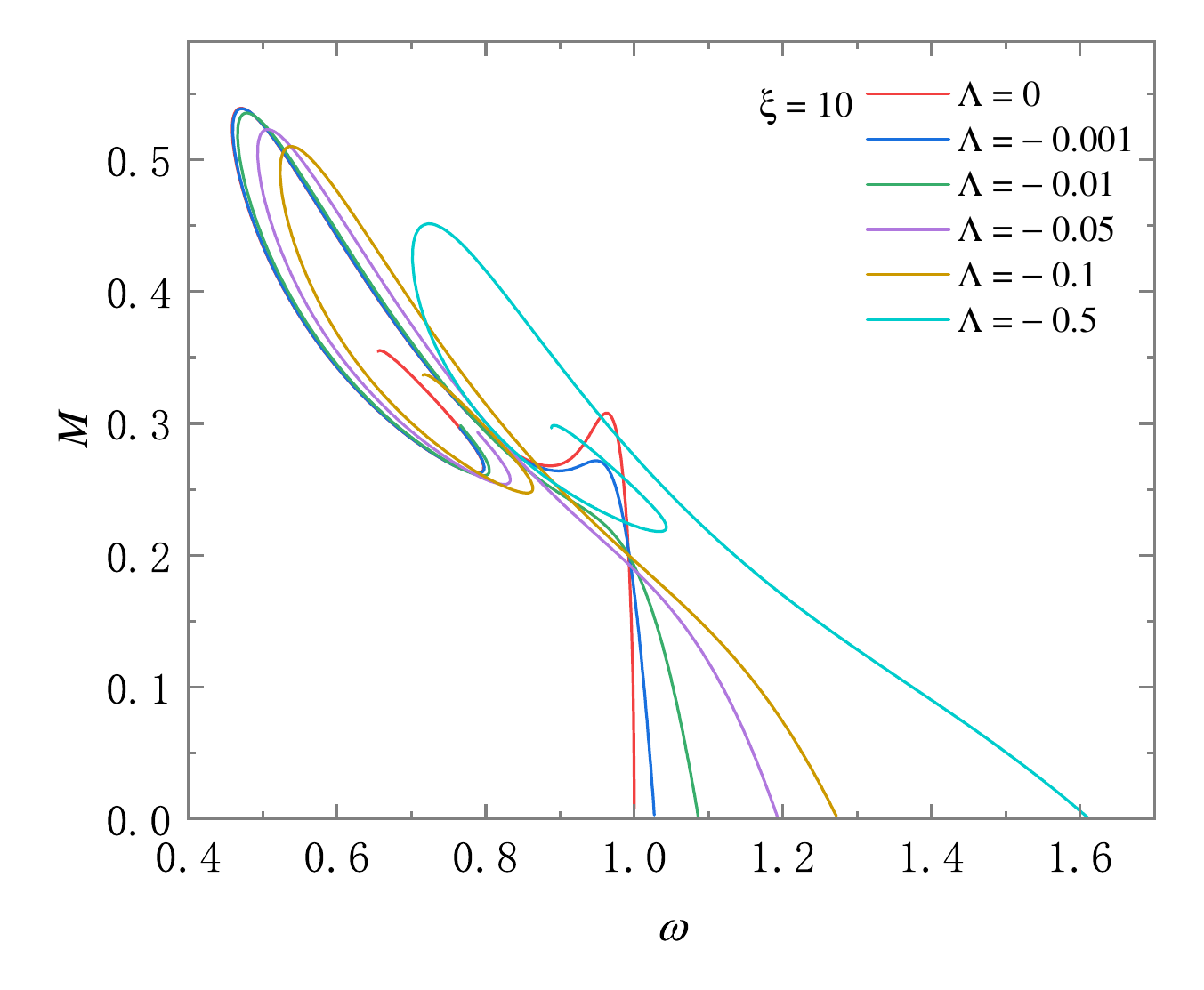}}
         \subfigure{\includegraphics[width=0.49\textwidth]{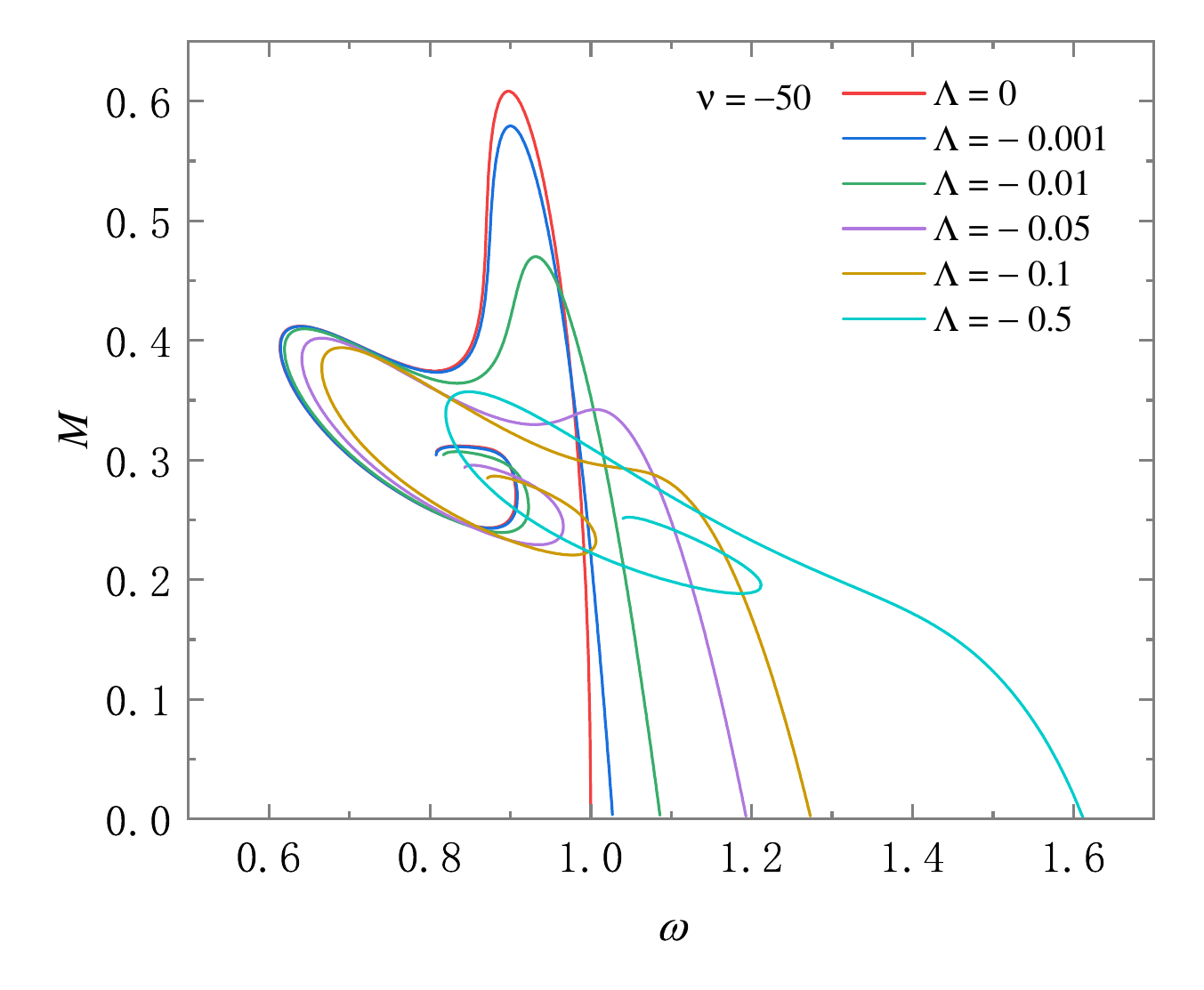}}
         \subfigure{\includegraphics[width=0.49\textwidth]{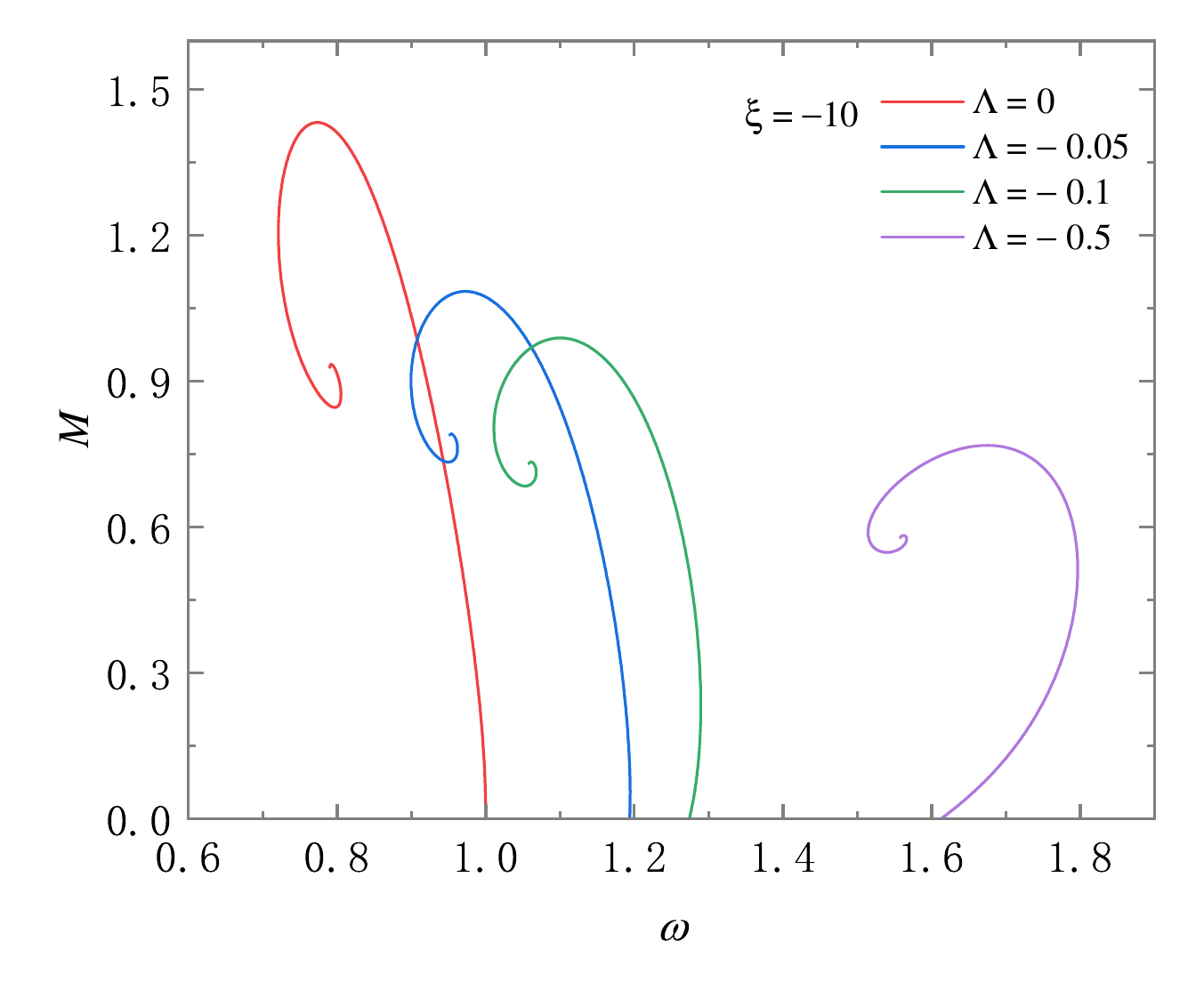}}
         \subfigure{\includegraphics[width=0.49\textwidth]{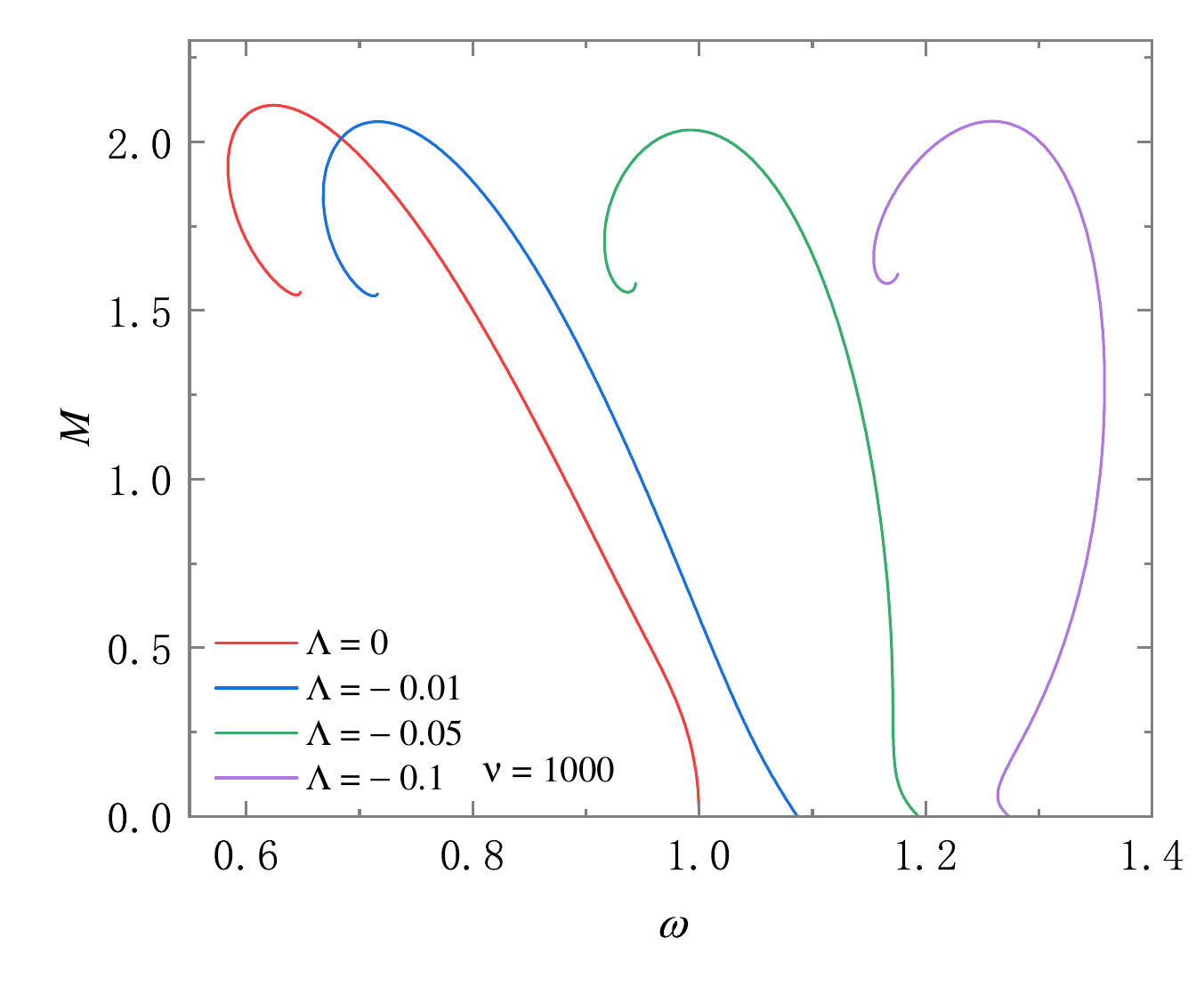}}
         \caption{The mass $M$ of the Dirac stars with different interactions as a function of frequency $\omega$ under different cosmological constants $\Lambda$. The left column corresponds to $\Phi^{4}$ interaction, while the right column corresponds to $\Phi^{6}$ interaction.}
         \label{psiMQ}
    \end{figure}

In Fig. \ref{psi46MQ}, we show the mass $M$ and charge $Q$ of the Dirac stars versus frequency $\omega$ for two different cosmological constants $\Lambda$ and various interaction parameters $\xi$ or $\nu$. The top two panels represent the Dirac stars in asymptotically Minkowski spacetime ($\Lambda=0$), while the bottom two panels illustrate the Dirac stars in AdS spacetime ($\Lambda=-0.1$). For positive interactions ($\xi<0$ or $\nu>0$), we observe that both $M_{max}$ and $Q_{max}$ increase gradually  as the absolute value of $\xi,\nu$ increase. For negative interactions ($\xi>0$ or $\nu<0$), at $\Lambda=0$, the mass and charge of the Dirac stars exhibit two peaks as the absolute values of $\xi$ and $\nu$ increase. For instance, for $\xi=30$ and $\nu=-50$,  it is evident that there are pronounced peaks at higher frequencies. At $\Lambda=-0.1$, increasing $\xi$ does not lead to the appearance of two peaks. Instead, as the frequency increases, both the mass and charge gradually approach zero monotonically from their respective maximum values. In the case of $\xi=-10$, the first branch exhibits a bulge, with the maximum frequency exceeding the frequency $\omega_{\textrm{v}}$. This property is associated with the cosmological constant and we show it in the bottom-left panel of Fig. \ref{psiMQ}. Unlike the quartic interaction, the sextic interaction, with decreasing $\nu$ shows a small fluctuation and exhibits two peaks. Whether in asymptotically Minkowski spacetime or AdS spacetime, interactions also alter the position and size of the spirals. However, they do not affect the value of $\omega_{\textrm{v}}$.
    \begin{figure}[h]
        \centering
        \subfigure{\includegraphics[width=0.49\textwidth]{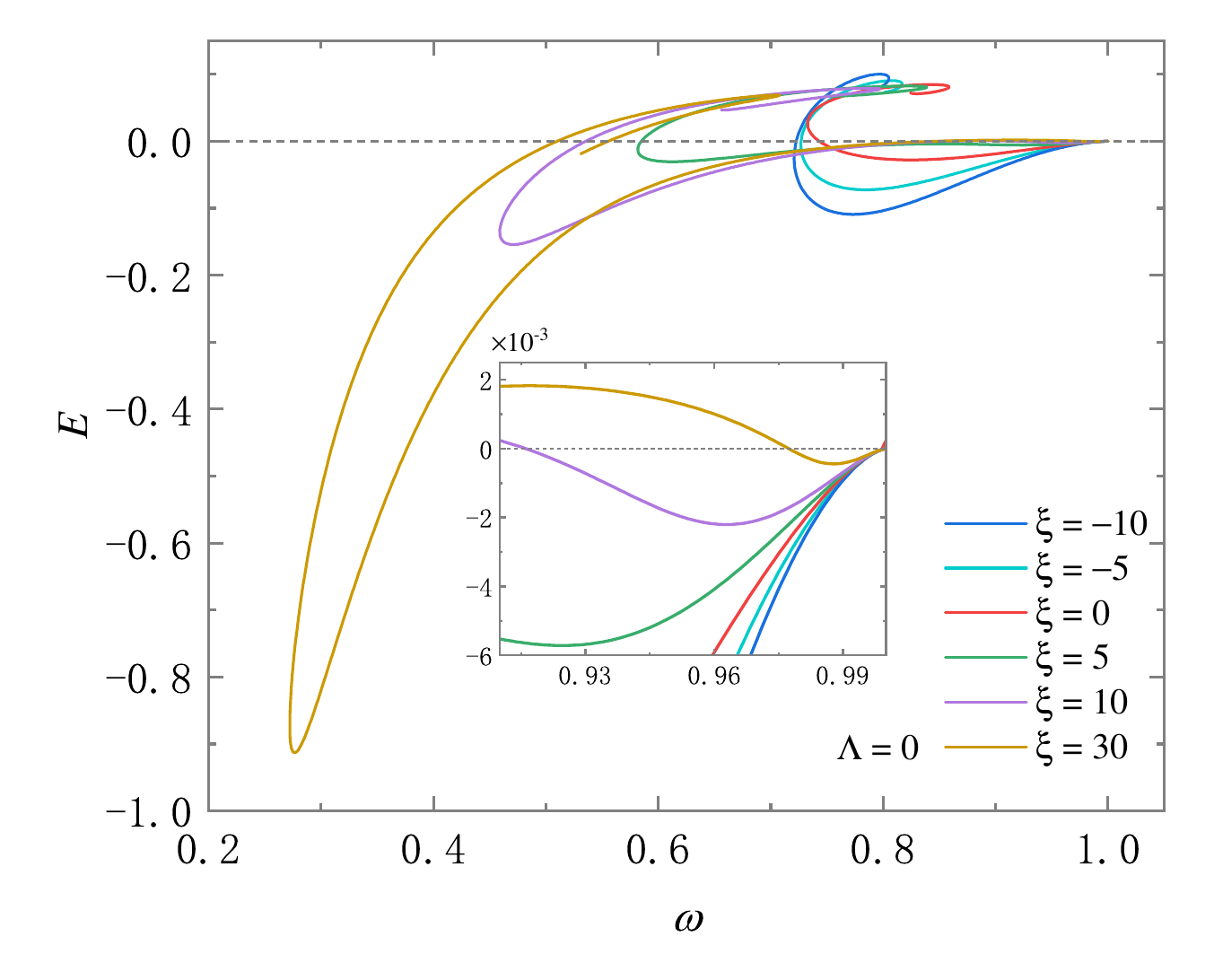}}
        \subfigure{\includegraphics[width=0.49\textwidth]{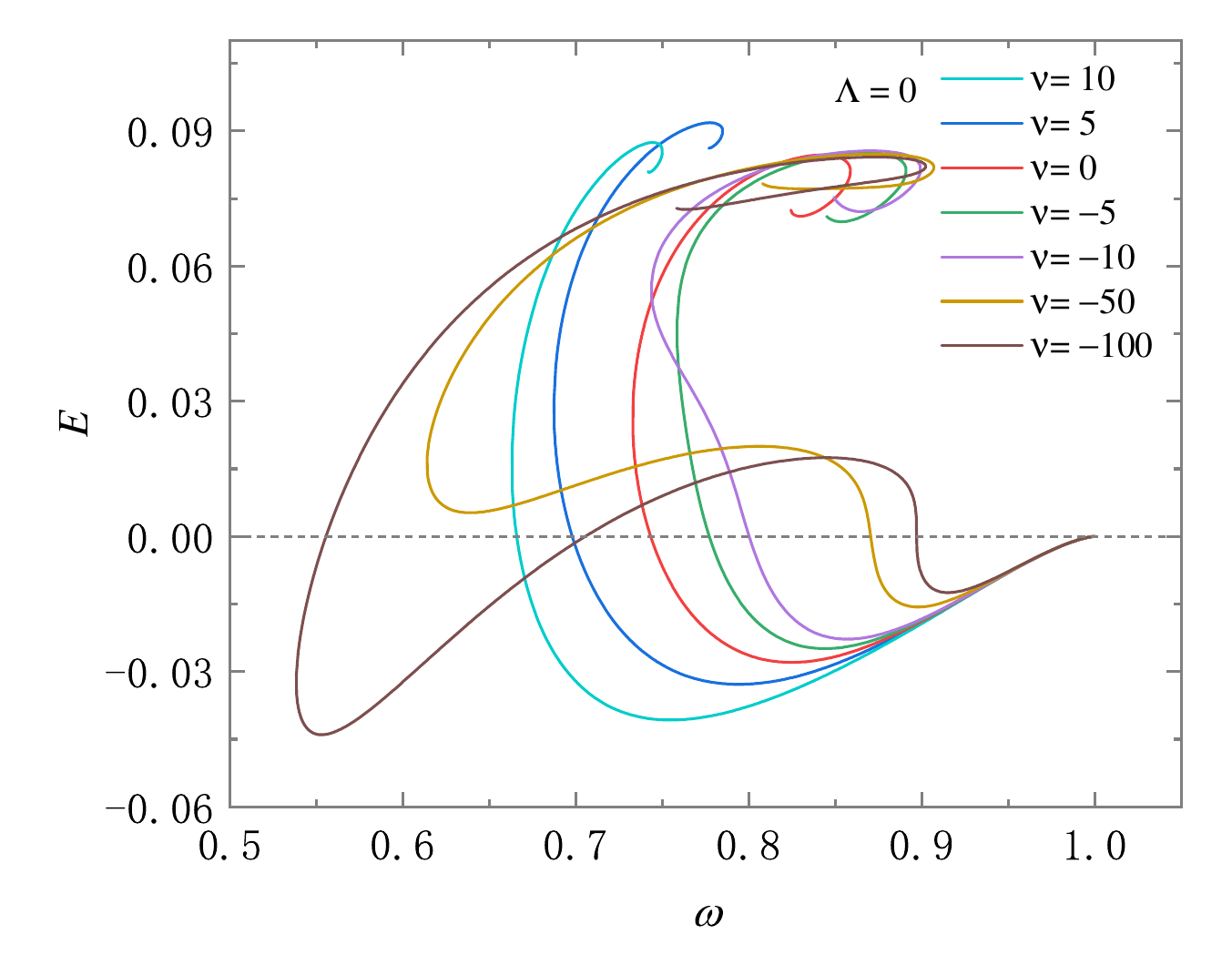}}
        \subfigure{\includegraphics[width=0.49\textwidth]{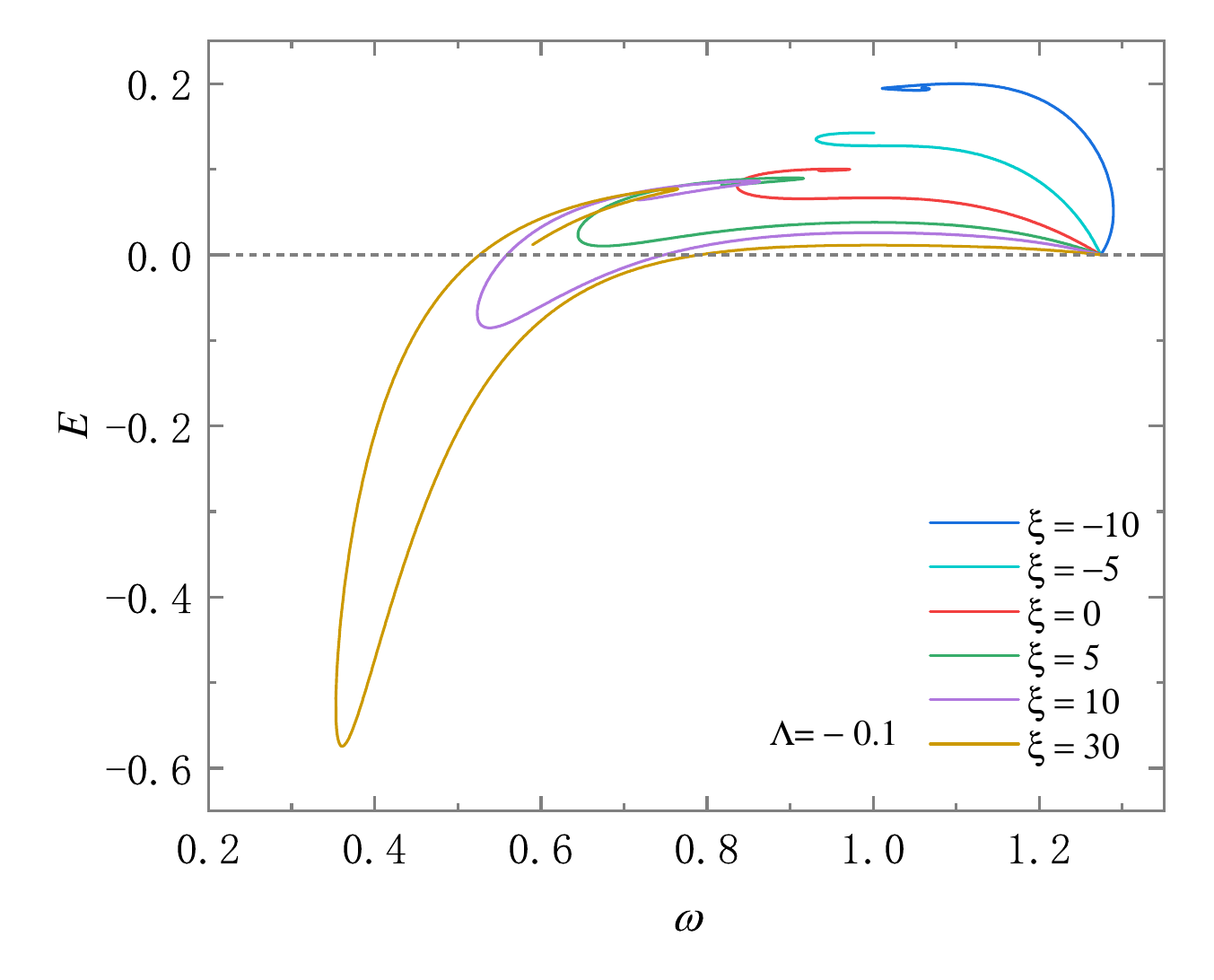}}
        \subfigure{\includegraphics[width=0.49\textwidth]{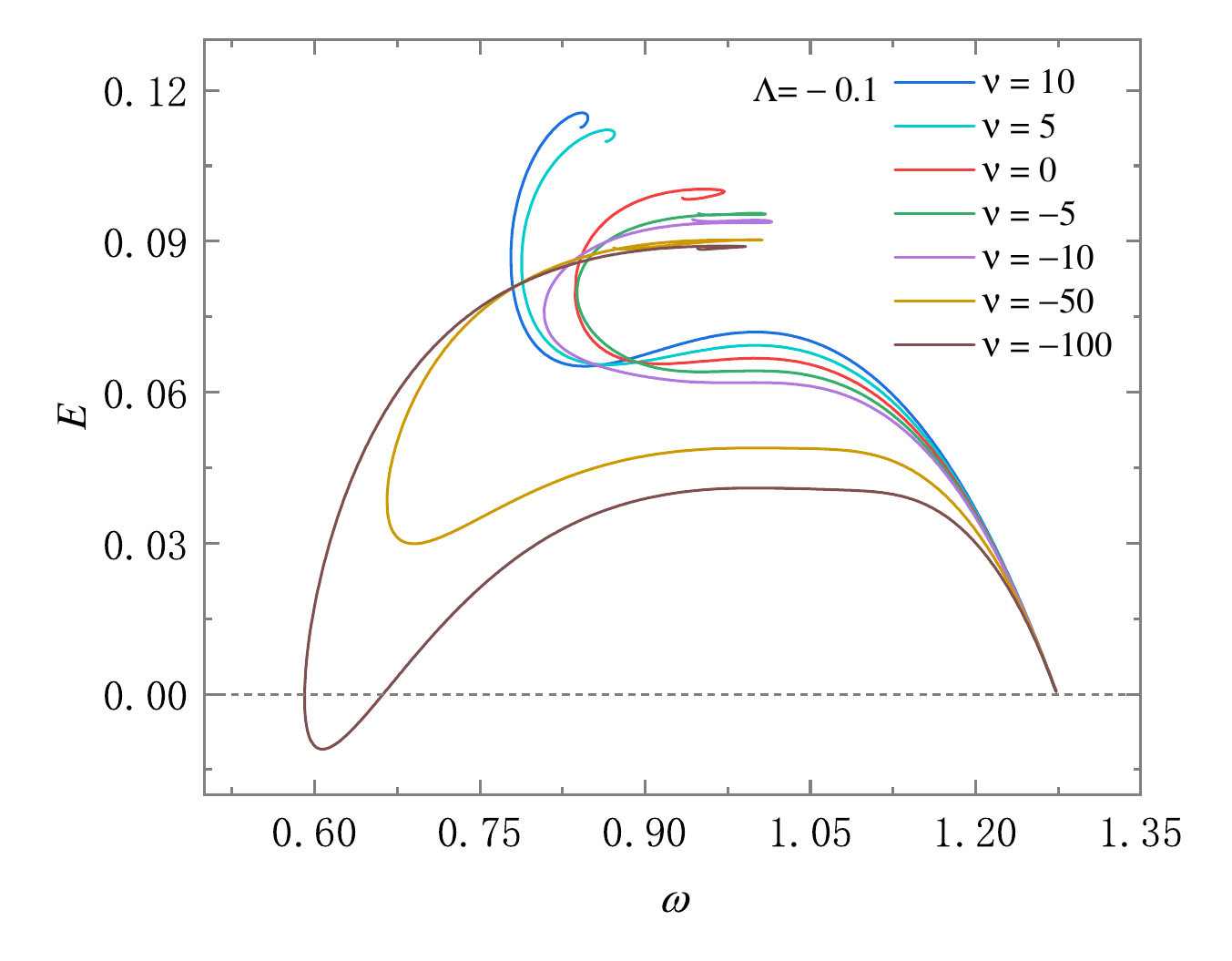}}
        \caption{The binding energy $E$ of the Dirac stars as a function of frequency $\omega$ for different interaction parameters $\xi$ or $\nu$ in asymptotically Minkowski spacetime ($\Lambda=0$) or AdS spacetime ($\Lambda=-0.1$), with the inset showing a detailed view at $\omega_ {\textrm{v}}$. The left column corresponds to $\Phi^{4}$ interaction, while the right column corresponds to $\Phi^{6}$ interaction.}
        \label{psi46E}
    \end{figure}

The energy binding $E$ as a function of the frequency $\omega$ for different interaction parameters $\xi$ or $\nu$ is illustrated in Fig. \ref{psi46E}.  In the case of $\Lambda=0$ (top two panels), both quartic and sextic interactions result in stable solutions for the Dirac stars. As the absolute value of $\xi$ increases, the minimum binding energy $E_{min}$ of the Dirac stars gradually decreases. The inset in the top-left panel depicts details at $\omega_{\textrm{v}}$, two peaks emerge as $\xi$ increases, and indicating the presence of a small stable region. On the other hand, when $\nu$ decreases to a certain value, and two minima in binding energy appear. Further decreasing $\nu$ causes both minima to become negative, and there exists a range of frequencies below $\omega_{\textrm{v}}$ where consistently exists negative binding energy. In the case of $\Lambda=-0.1$ (bottom two panels), increasing $\xi$ results in stable solutions for the Dirac stars, and greater $\xi$ values correspond to smaller $E_{min}$. As $\nu$ decreases, stable solutions also emerge for the Dirac stars. In AdS spacetime, with an attractive interaction ($\xi>0$ or $\nu<0$), the Dirac stars become more stable.

\begin{figure}[h]
        \centering
        \subfigure{\includegraphics[width=0.49\textwidth]{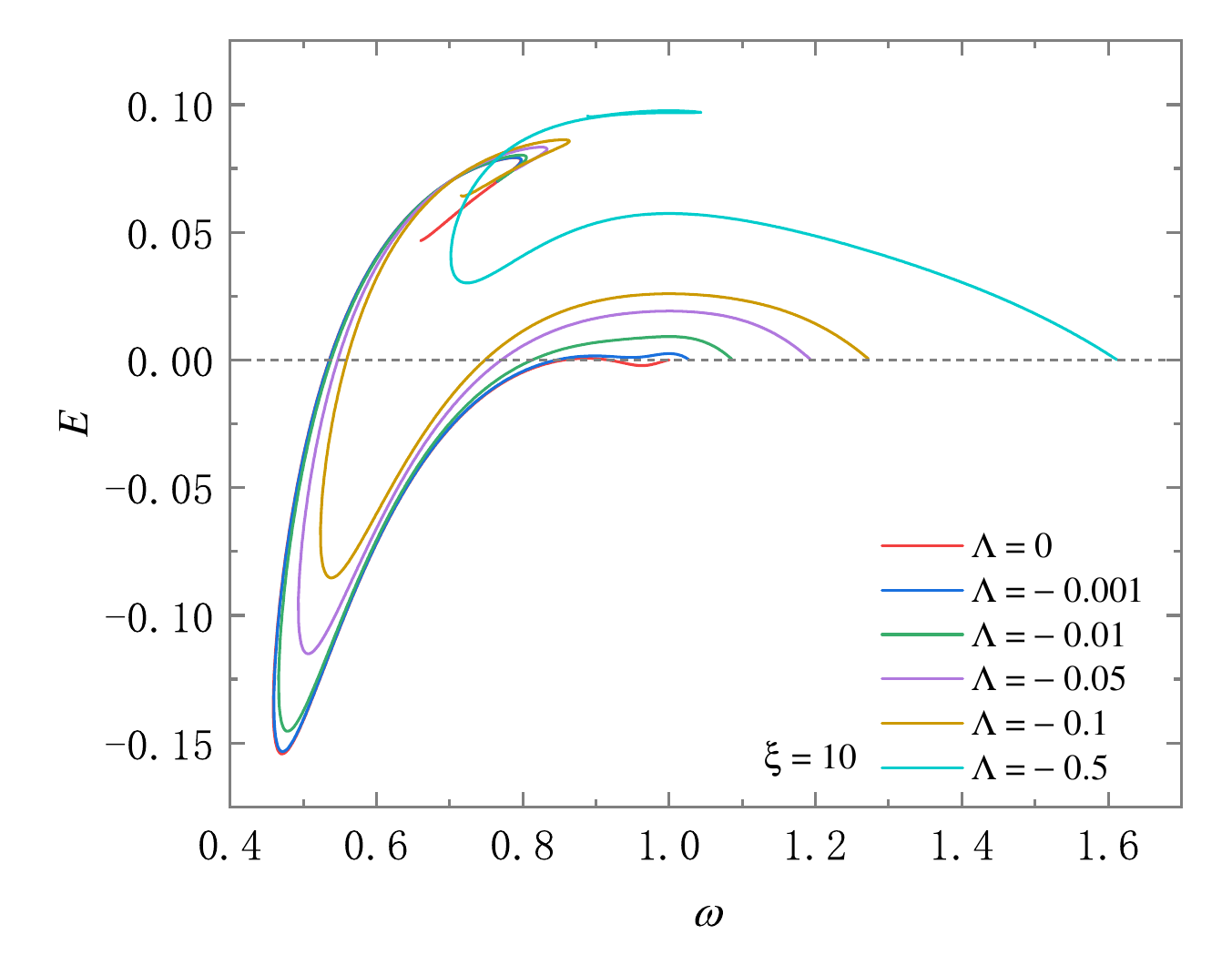}}
        \subfigure{\includegraphics[width=0.49\textwidth]{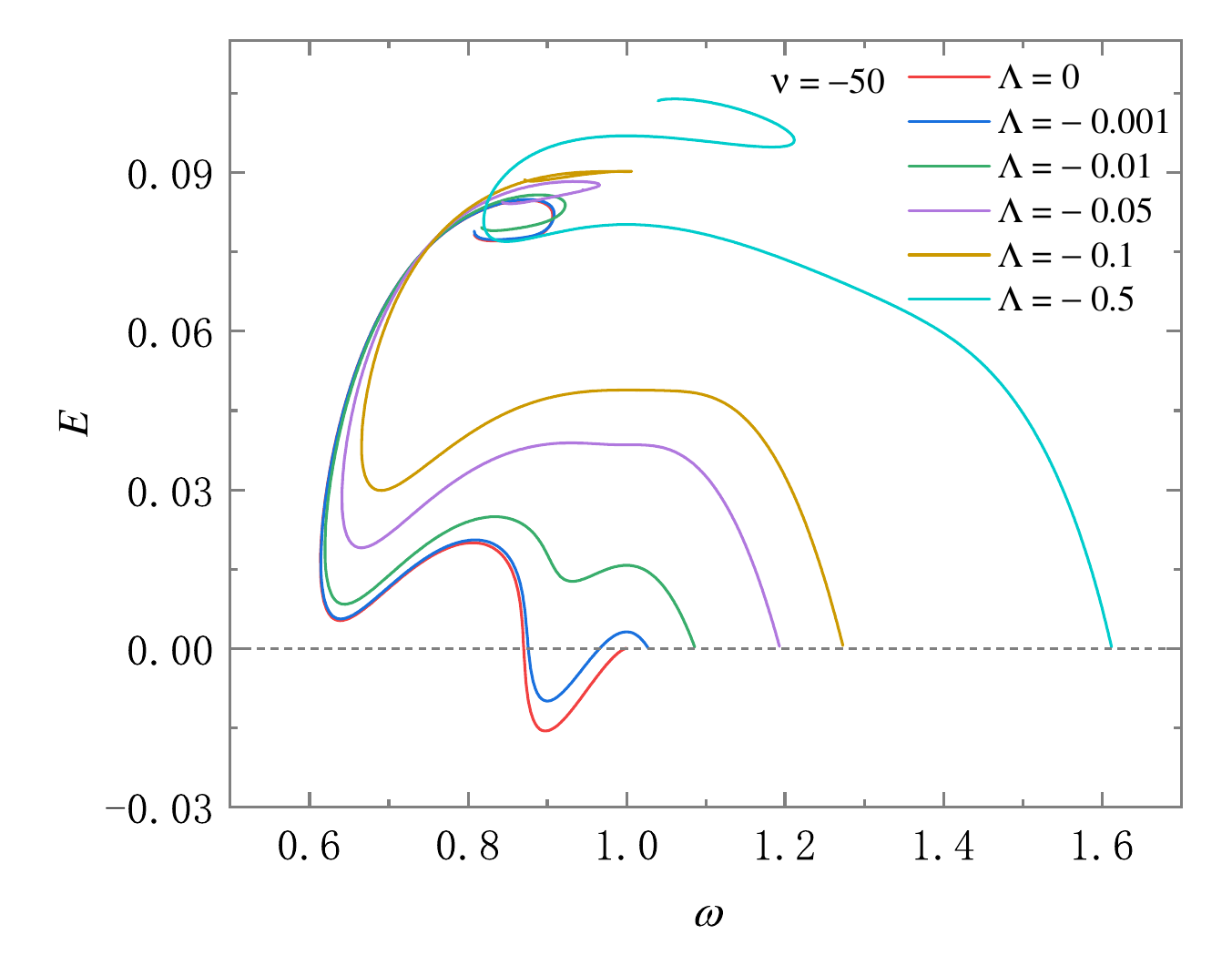}}
        \subfigure{\includegraphics[width=0.49\textwidth]{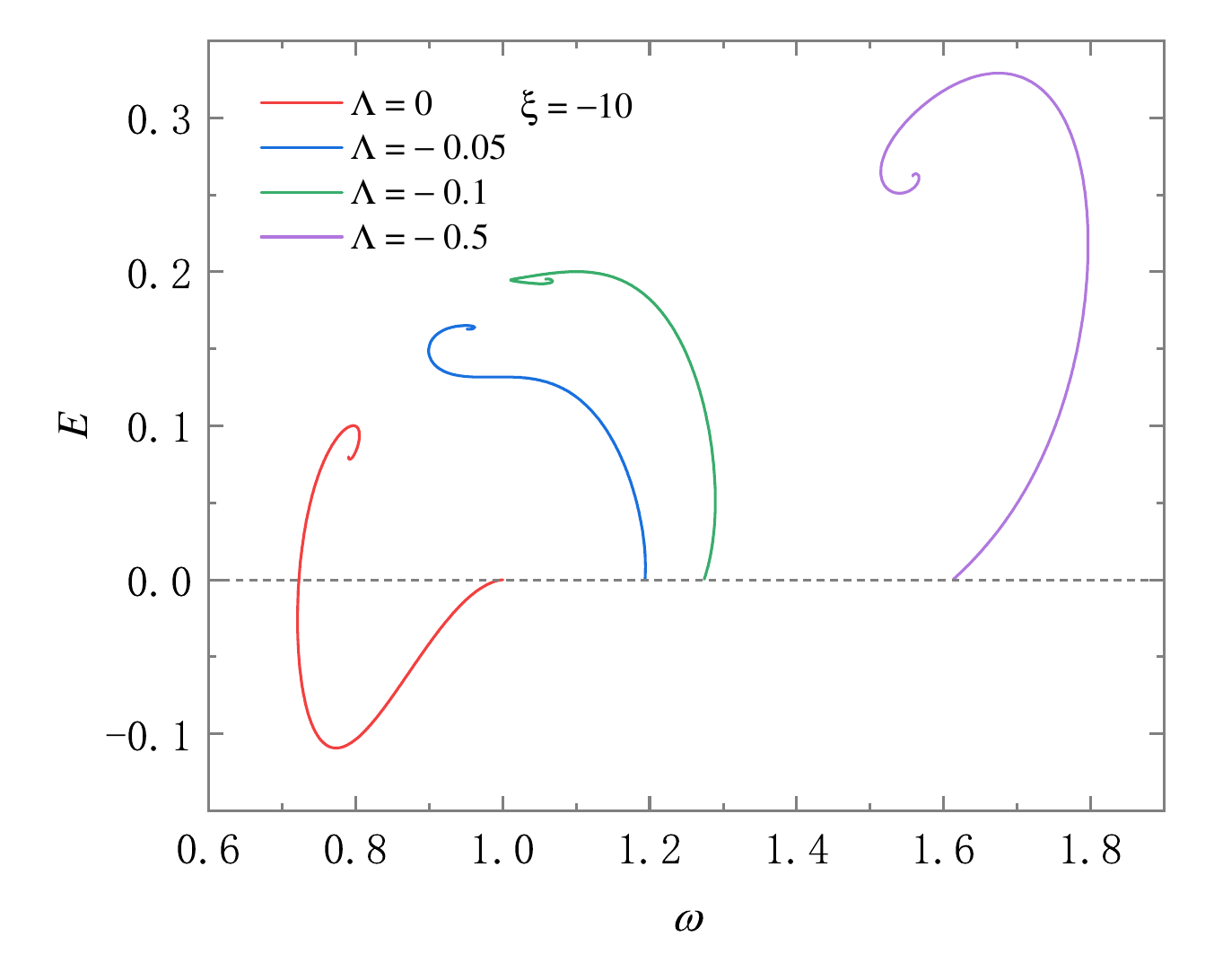}}
        \subfigure{\includegraphics[width=0.49\textwidth]{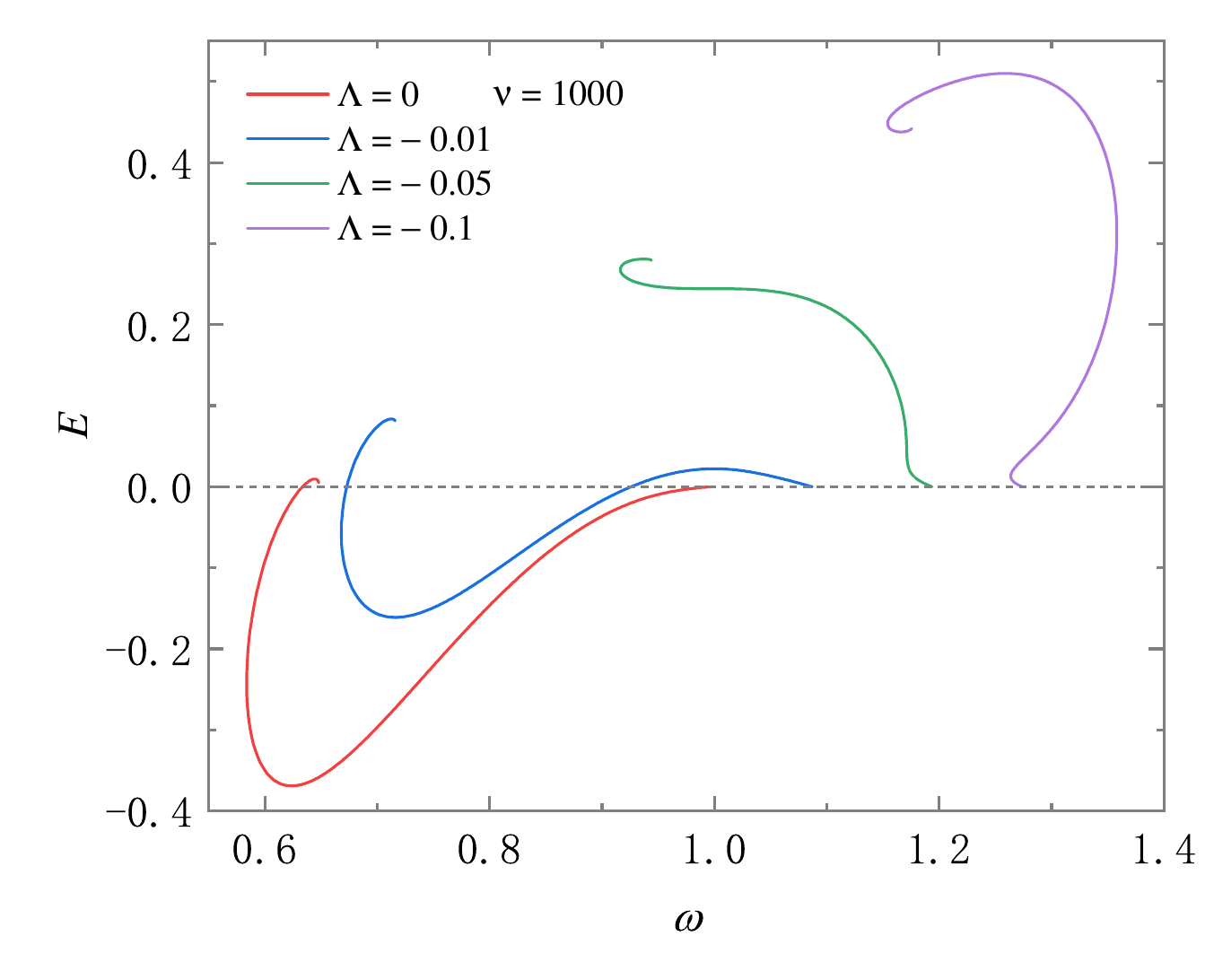}}
        \caption{The  binding energy $E$ of the Dirac stars with different interactions as a function of frequency $\omega$ for different cosmological constants $\Lambda$. The left column corresponds to the $\Phi^{4}$ interaction, while the right column corresponds to the $\Phi^{6}$ interaction.}
        \label{psiE}
    \end{figure}

To study the influence of the cosmological constant on the Dirac stars with quartic or sextic interactions more clearly, we have chosen four different interaction parameters to plot Fig. \ref{psiMQ} and Fig. \ref{psiE}. We investigate the relationship between the mass $M$ of the Dirac stars with different cosmological constants $\Lambda$ and the frequency $\omega$ in Fig. \ref{psiMQ} . For negative interactions (top two panels), we observe that the distribution of the spiral position is more dense with the change of the cosmological constant, and the appearance of two maxima gradually disappears as the cosmological constant decreases. For positive interactions (bottom two panels), the spiral position is more scattered with the change of the cosmological constant. At $\xi=-10$ and $\Lambda=-0.5$, it is evident that the spiral extends beyond $\omega_{\textrm{v}}$, with $\omega_{max}>\omega_{\textrm{v}}$. For comparison with quartic interactions, we take $\nu=1000$, and also observe the spiral extending beyond $\omega_{\textrm{v}}$. However, the variation of mass with frequency is not necessarily monotonic near $\omega_{\textrm{v}}$. For example, at $\Lambda=-0.1$, the frequency initially decreases as the mass of the Dirac stars increases. Then, the frequency increases, reaching a maximum frequency $\omega_{max}$, and finally forming a spiral.

Fig. \ref{psiE} shows the variation of binding energy $E$ of the Dirac stars with frequency $\omega$ under different cosmological constants $\Lambda$. When the cosmological constant is relatively small, the Dirac stars exhibit stable solutions. As the cosmological constant decreases further, the Dirac stars gradually become unstable. In the case of negative interactions (top two panels), the binding energy of the Dirac stars with $\Lambda=-0.5$ does not form a spiral. It exhibits the binding energy on the third branch is higher than that on the second branch, and binding energy on the second branch is higher than that on the first branch. In the case of positive interactions (bottom two panels), as the cosmological constant decreases, the spiral direction of the binding energy of the Dirac stars gradually reverses.

\subsection*{Case 3: $\xi=1$ and $\nu=1$}
\begin{figure}[h]
        \centering
        \subfigure{\includegraphics[width=0.49\textwidth]{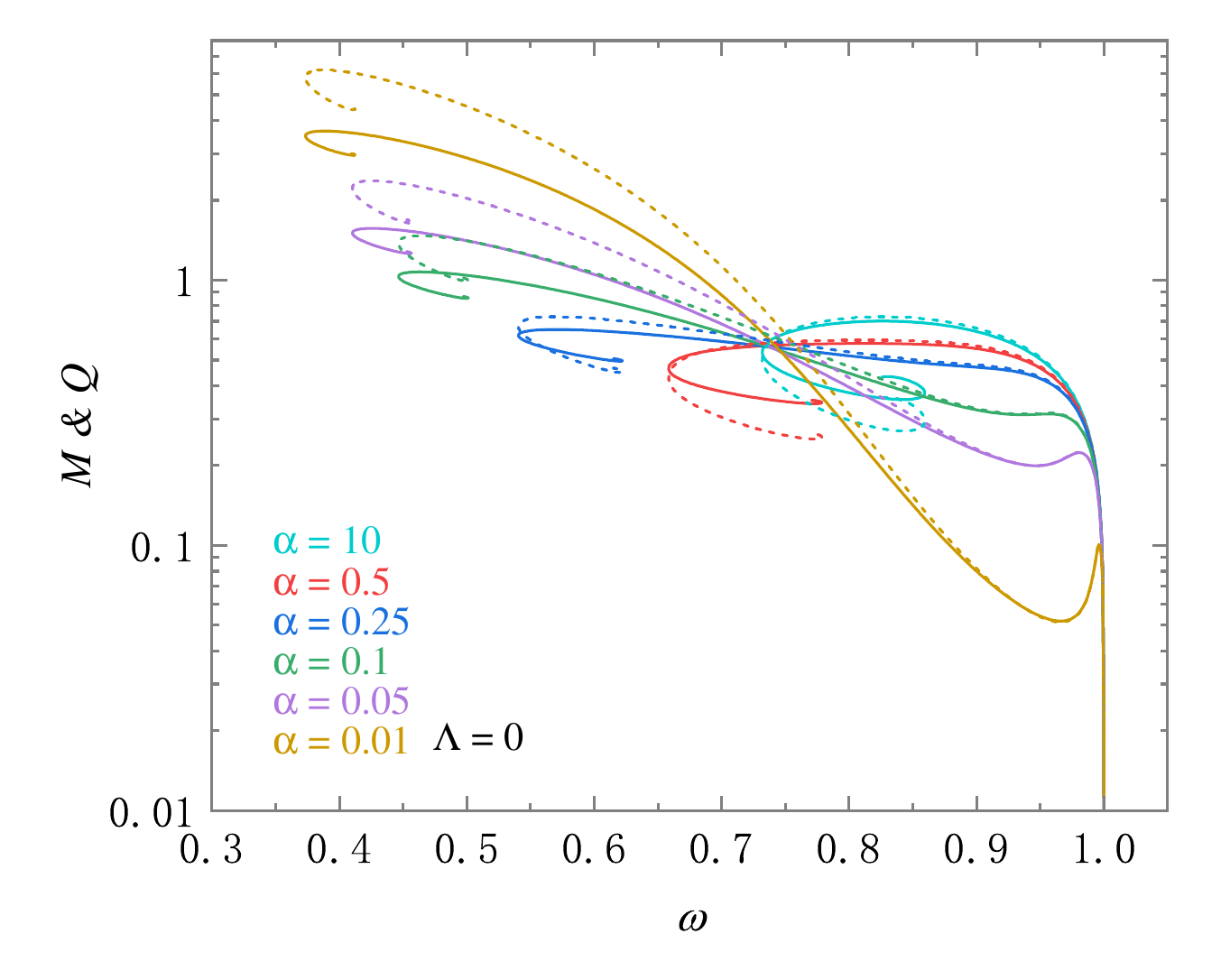}}
        \subfigure{\includegraphics[width=0.49\textwidth]{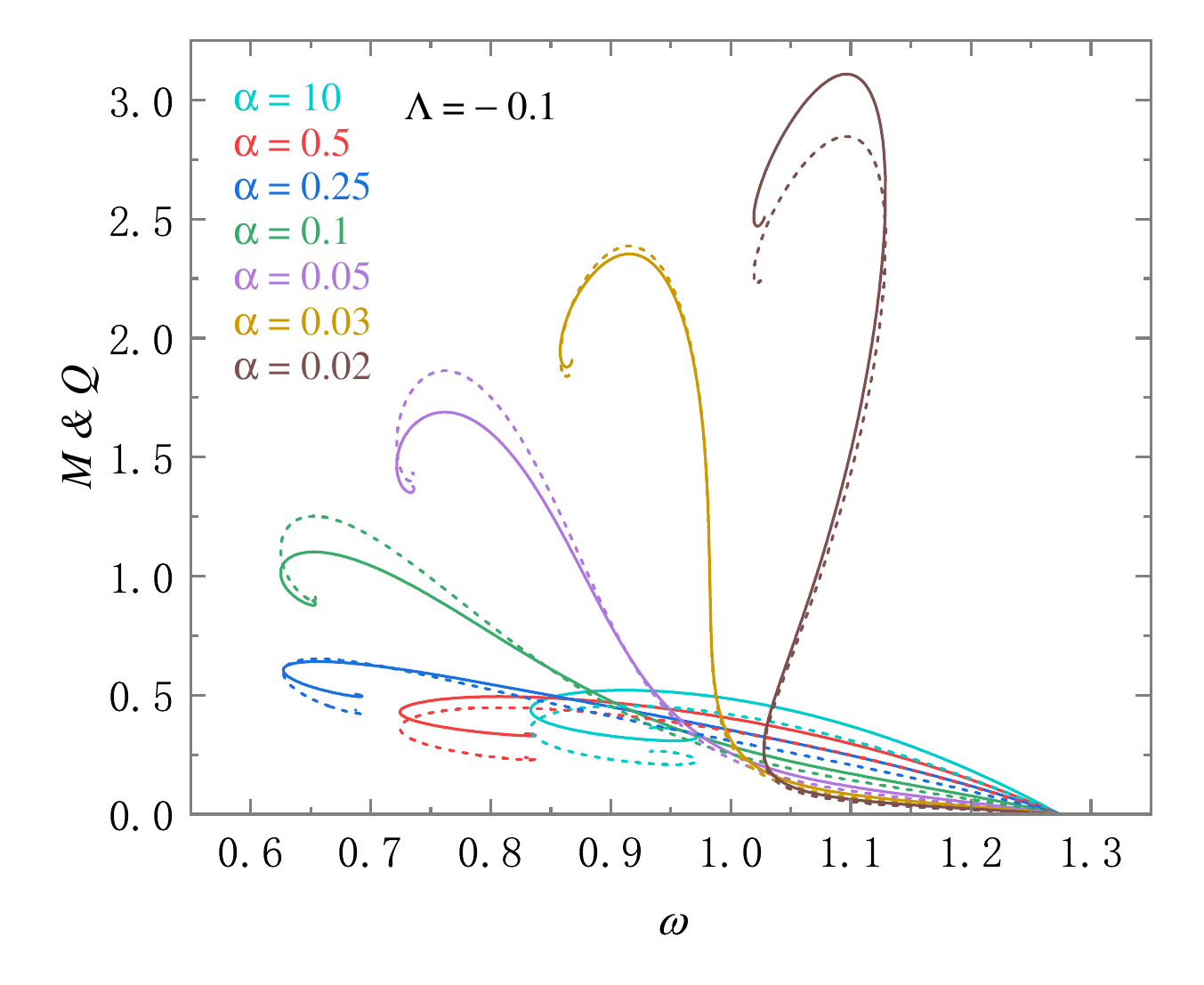}}
        \caption{The mass $M$ (solid line) and the charge $Q$ (dashed line) of the Dirac stars with different parameters $\alpha$ as functions of frequency $\omega$. Left: asymptotically Minkowski spacetime ($\Lambda=0$). Right: AdS spacetime ($\Lambda=-0.1$).}
        \label{starMQ}
    \end{figure}

In this section, we consider the effect of Q-ball type interactions with fixed interaction coupling constants $\xi=1$ and $\nu=1$. Fig. \ref{starMQ} shows the variation of mass $M$ and charge $Q$ with frequency $\omega$ for different values of the parameter $\alpha$. When $\Lambda=0$ (left panel), we observe that the mass and charge of the Dirac stars exhibit two peaks as $\alpha$ decreases. When $\Lambda=-0.1$ (right panel), the high-frequency peak disappears. Due to the influence of the cosmological constant, as $\alpha$ decreases, the mass and charge of the Dirac stars on first branch divide into three small branches. For example, in the curve with $\Lambda=-0.1$ and $\alpha=0.02$, mass and charge start increasing from zero as the frequency decreases on the first small branch. Then, they increase with increasing frequency on the second small branch. Finally, on the third small branch, mass and charge reach $M_{\max}(Q_{\max})$ as the frequency decreases again, and connecting back to the end of the first branch.

\begin{figure}
        \centering
        \subfigure{\includegraphics[width=0.49\textwidth]{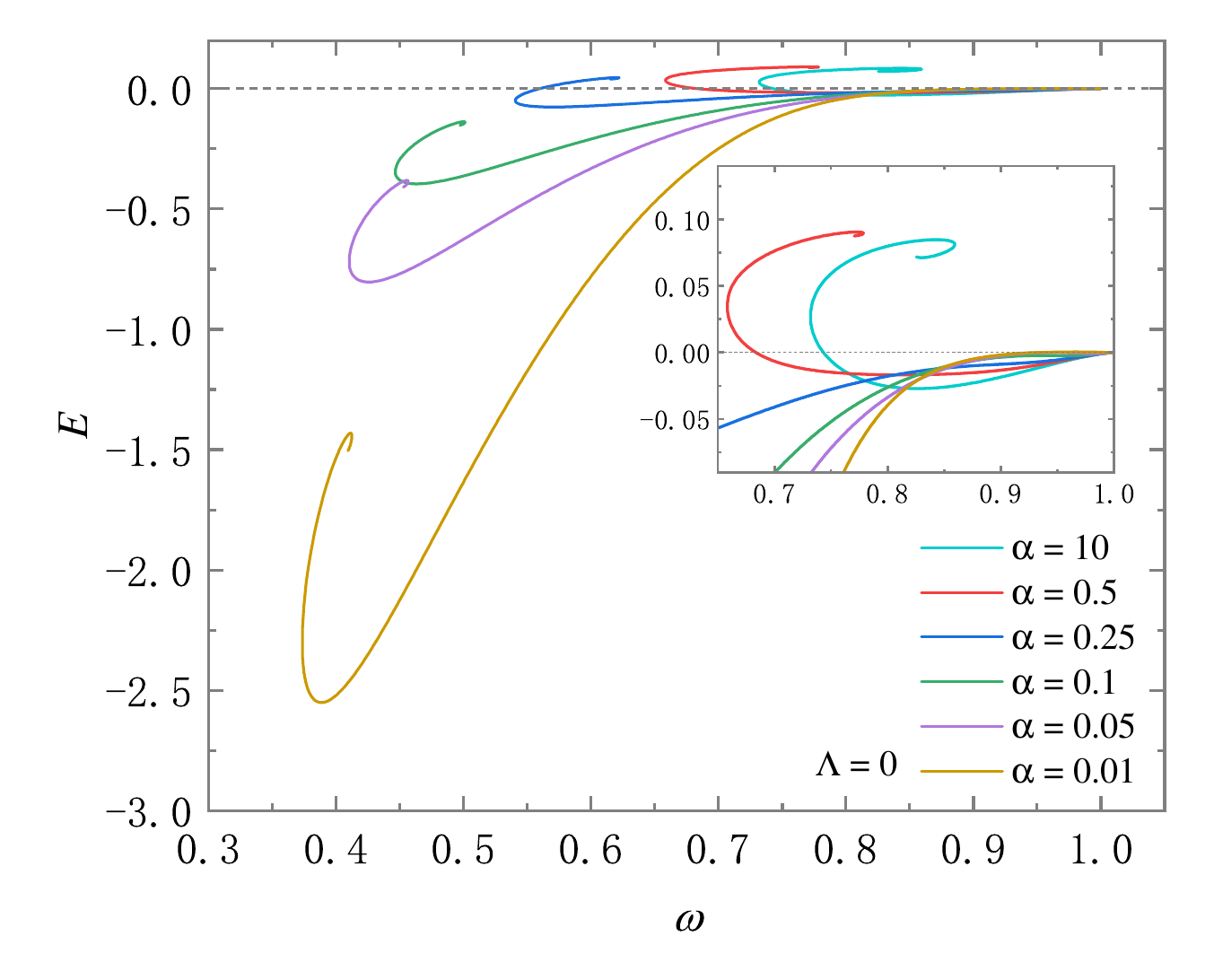}}
        \subfigure{\includegraphics[width=0.49\textwidth]{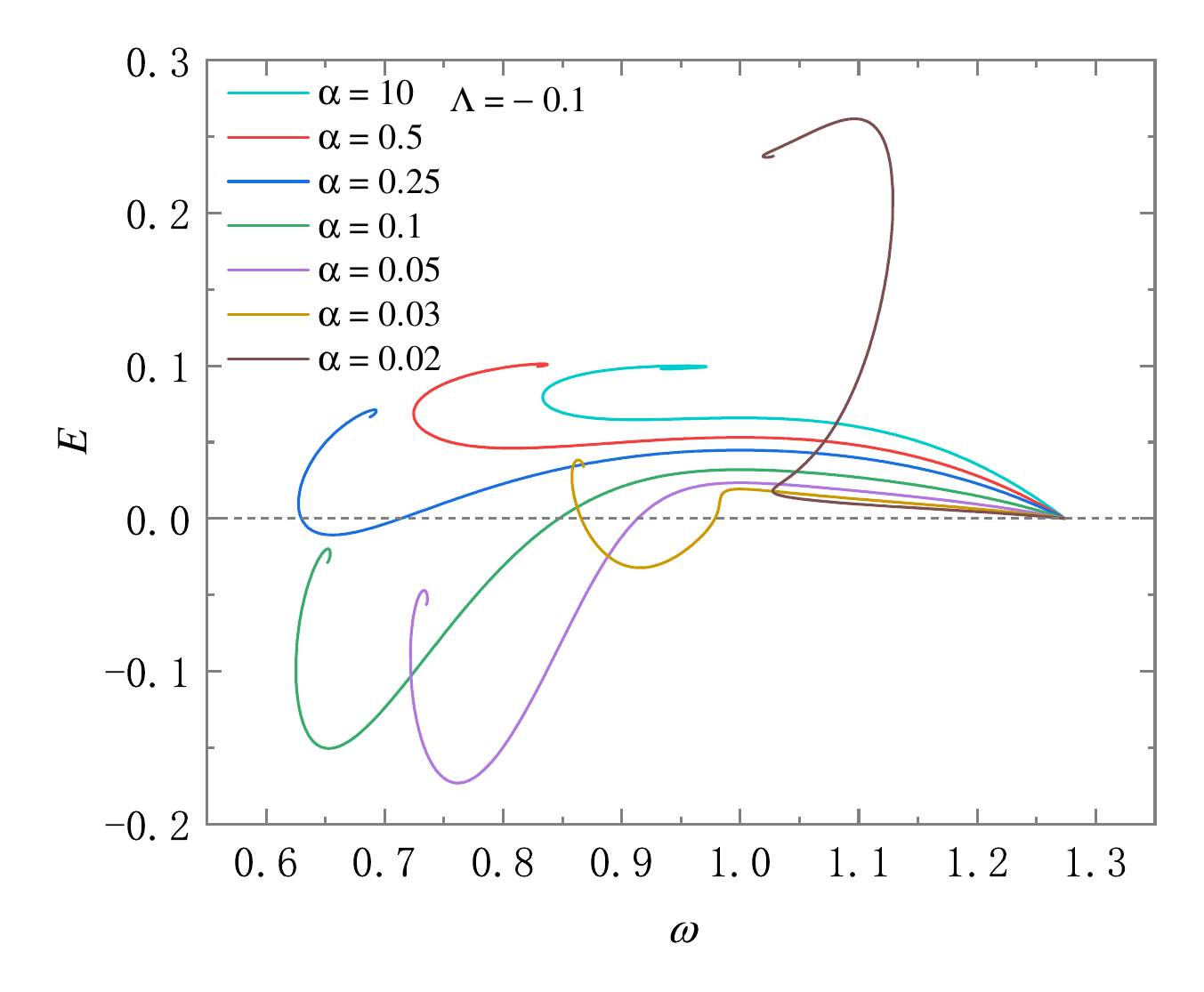}}
        \caption{The binding energy $E$ of the Dirac stars with different parameters $\alpha$ as a function of frequency $\omega$, with the inset detailing the behavior at $\omega=1$. Left: asymptotically Minkowski spacetime ($\Lambda=0$). Right: AdS spacetime ($\Lambda=-0.1$).}
         \label{starE}
    \end{figure}

In Fig. \ref{starE}, we demonstrate the binding energy $E$ of the Dirac stars with Q-ball type interactions at different $\alpha$ as a function of frequency $\omega$. When $\Lambda=0$ (left panel), we illustrate the details at $\omega=1$ in the inset, and it can be observed that stable solutions exist for various $\alpha$. Moreover, $E_{min}$ of the Dirac stars decreases as $\alpha$ decreases, which indicates increased stability. When $\Lambda=-0.1$ (right panel), stable solutions are limited to a certain range of $\alpha$, and $\alpha$ also influences the spiral direction of the binding energy.

\begin{figure}[b]
        \centering
        \subfigure{\includegraphics[width=0.49\textwidth]{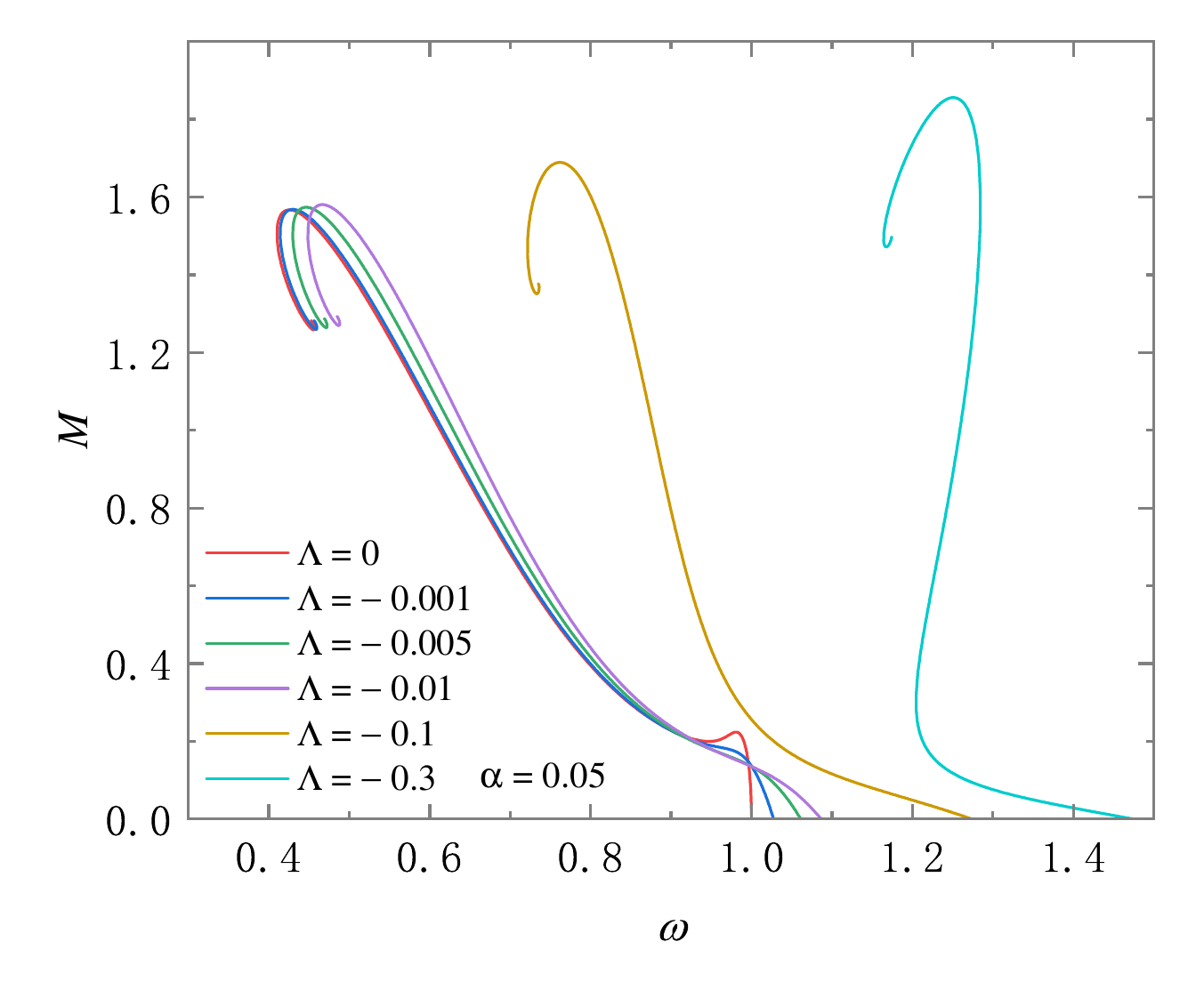}}
        \subfigure{\includegraphics[width=0.49\textwidth]{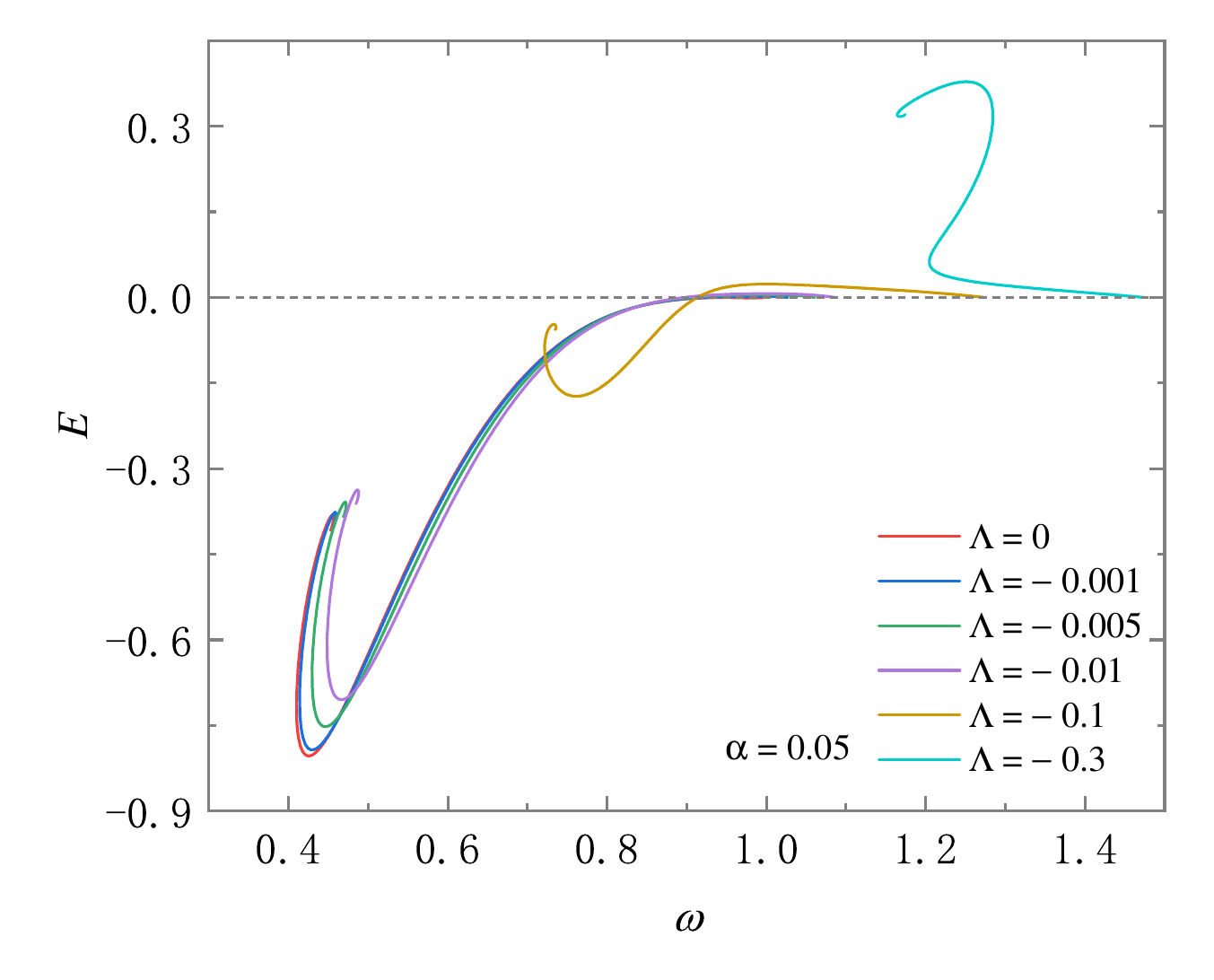}}
        \caption{The mass $M$ (left) and binding energy $E$ (right) as functions of frequency $\omega$ for the Dirac stars at different cosmological constants $\Lambda$, with $\alpha=0.05$.}
        \label{starc}
    \end{figure}

Similarly, to further investigate the impact of the cosmological constant on the Dirac stars with Q-ball type interactions, we depict Fig. \ref{starc}. It presents the variations in mass $M$ (left) and binding energy $E$ (right) with frequency$\omega$ for the Dirac stars ($\alpha=0.05$) at different cosmological constants $\Lambda$. We observe that the peak of the Dirac stars at large frequencies gradually disappears as the cosmological constant decreases. Differing from the behavior discussed earlier for the Dirac stars, both $M_{max}$ and $Q_{max}$ increase as the cosmological constant decreases. The Dirac stars gradually has no stable solutions with the decrease of the cosmological constant.

     \begin{figure}
        \centering
        \subfigure{\includegraphics[width=0.49\textwidth]{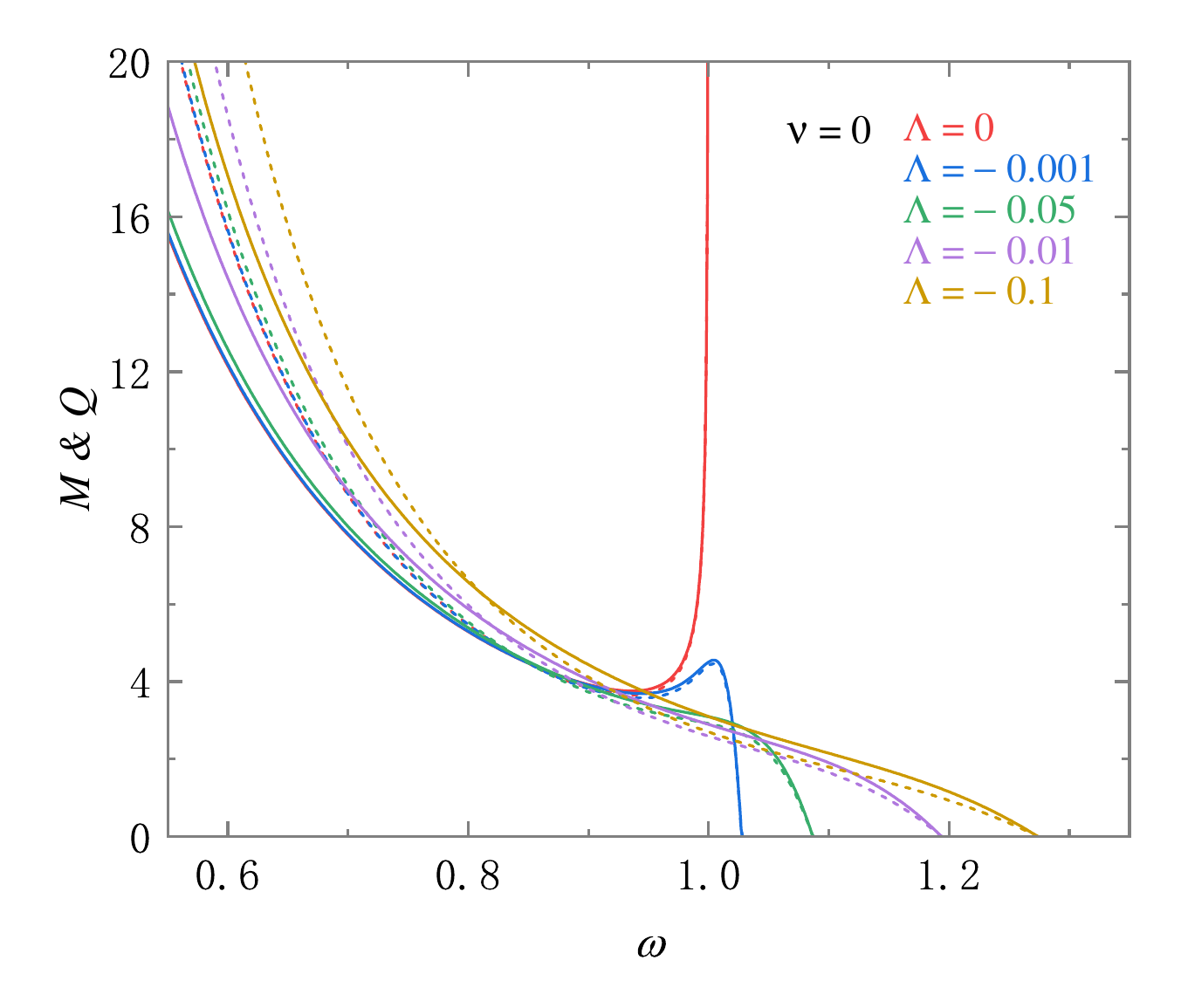}}
        \subfigure{\includegraphics[width=0.49\textwidth]{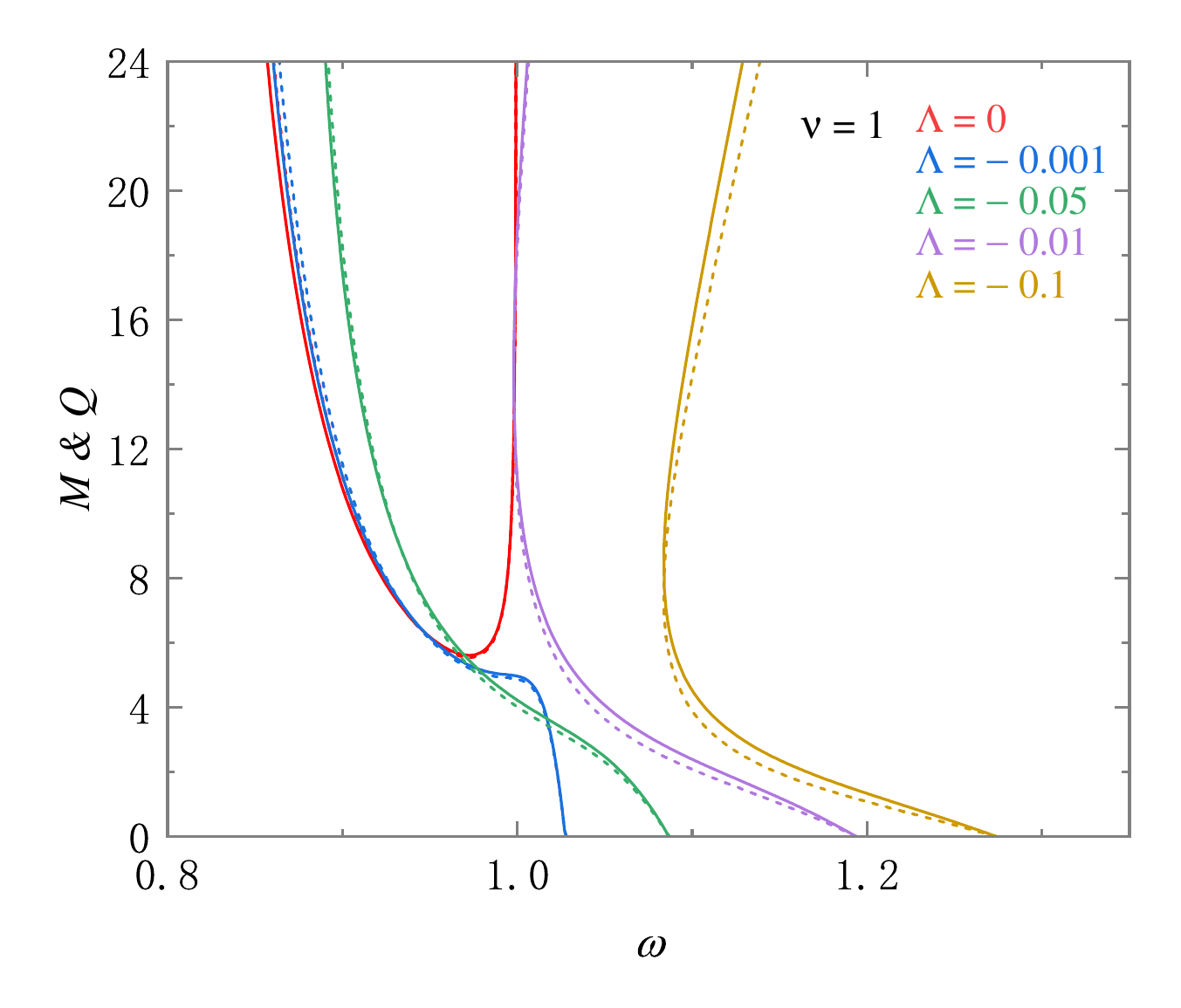}}
        \caption{The mass $M$ (solid line) and charge $Q$ (dashed line) of Dirac stars as functions of frequency $\omega$ for different cosmological constants $\Lambda$. Left: only quartic interaction ($\xi=1,\nu=0$). Right: Q-ball type interaction ($\xi=1,\nu=1$).}
        \label{picflat}
     \end{figure}
We supplement the situation of the Dirac Q-balls, which decouple from the interacting Dirac stars. We fix $\alpha=0$, resulting in $m(r)=0$, $\sigma(r)=1$, and consequently $N(r)=1-\Lambda r^{2}/3$. In this situation, we only need to solve the Dirac field equations (\ref{eom12})(\ref{eom13}), with the initial and boundary conditions the same as those for the Dirac stars. In Fig. \ref{picflat}, we examine the variations of mass $M$ and charge $Q$ with frequency $\omega$ under different cosmological constants $\Lambda$. The left panel corresponds to the case with only quartic interactions ($\nu=0$), while the right panel represents the Q-ball type interactions case ($\nu=1$). In previous work \cite{Herdeiro:2020jzx}, the Dirac Q-balls in Minkowski background were studied, corresponding to the red lines in our panels. It can be observed that both mass and charge exist in a certain frequency range, $\omega_{min}<\omega<\omega_{max}=\mu$, where they tend to diverge at $\omega_{min}$ and $\omega_{max}$ respectively. When the cosmological constant is nonzero, the mass and charge tend to zero at a specific frequency $\omega_{\textrm{v}}$. This behavior is similar to the Dirac stars with gravity, and $\omega_{\textrm{v}}$ varies with the cosmological constant just like in the case of the Dirac stars. Due to the constraints of the AdS background, the integral is finite for $\Lambda\neq0$ at one end and approaches infinity at the other, even if $\Lambda\neq0$. The behavior of Q-balls in the AdS background has been studied in \cite{Hartmann:2012gw}\cite{Hartmann:2012wa}, which is similar to the Dirac field model with quartic interactions. In a frequency range below a certain value, the charge is greater than the mass, $M-Q=E<0$, which ensures the existence of stable solutions. However, the situation is different for Q-ball type interactions. In the case of small cosmological constants, stable solutions exist within a smaller frequency range, as observed for $\Lambda=-0.001$ with stable solutions present below the frequency $\omega=0.93$. As the cosmological constant decreases further, stable solutions for the Dirac Q-balls do not appear. Additionally, the relationship between the mass, charge, and frequency for the Dirac Q-balls is non-monotonic. For $\Lambda=-0.1$, the frequency first decreases and then increases with an increase of mass and  charge.

\section{CONCLUSION}\label{sec5}
In this paper, we constructed the free field Dirac stars and self-interacting Dirac stars in AdS spacetime, as well as Dirac Q-balls in the AdS background. We discussed the ADM mass, charge, and binding energy of these solutions.

Firstly, we studied the free field Dirac stars. The spiral is a significant feature in the models we construct involving gravity, and the cosmological constant does not change this feature. However, the cosmological constant can alter the frequency $\omega_{\textrm{v}}$, which corresponds to the solutions being spatially diluted. Then, we extended the study to the self-interacting Dirac stars. For quartic or sextic interactions, the properties of the Dirac stars are quite similar. Positive interactions result in an increase in both $M_{max}$ and $Q_{max}$, while negative interactions result in two peaks for both mass and charge. These peaks gradually disappear as the cosmological constant decreases. The introduction of a cosmological constant leads to the gradual absence of stable solutions for the Dirac stars. Whether in asymptotically Minkowski or AdS spacetime, the appropriate interactions make the Dirac stars more stable. When considering both quartic and sextic interactions simultaneously, the Dirac stars also exhibit two peaks for mass and charge, and these peaks gradually disappear as the cosmological constant decreases. Furthermore, $M_{max}$ and $Q_{max}$ increase as the cosmological constant decreases. In asymptotically Minkowski spacetime, stable solutions for the Dirac stars exist regardless of the strength of gravitational effects, but this is not the case in AdS spacetime, where stable solutions are absent for weak gravitational effects. As a supplement, we also investigated Dirac Q-balls in the AdS background. In contrast to previous studies on Q-balls \cite{Hartmann:2012gw}\cite{Hartmann:2012wa} and Proca Q-balls \cite{Loginov:2015rya}, our research reveals that the Dirac Q-balls with quartic interactions have stable solutions. Moreover, for the Dirac Q-balls with Q-ball type interaction, the changes in mass and charge with frequency exhibit a non-monotonic behavior with the decrease of the cosmological constant.

As the parameter $\alpha$ varies from infinity to zero, the Dirac stars and the Dirac Q-balls correspond to each other. However, they exhibit significant differences in different spacetime and differ from models involving scalar and vector fields. This study only addressed the ground state of the Dirac field, but exploring excited states and charged Dirac stars is also part of our future work.

\section*{Acknowledgements}
This work is supported by the National Key Research and Development Program of China (Grant No. 2020YFC2201503) and the National Natural Science Foundation of China (Grant No.~12047501 and No.~12275110).

\end{document}